\documentclass[final,onefignum,onetabnum,onealgnum]{siamonline220329} 
\pdfoutput=1

\usepackage{amsfonts}
\usepackage{xcolor}
\usepackage{hyperref}
\usepackage{amsmath}
\usepackage{amssymb}
\usepackage{mathtools}
\usepackage{multirow}
\usepackage{enumitem}
\usepackage{graphicx}
\usepackage{xspace}
\usepackage[normalem]{ulem} 

\usepackage[noend]{algpseudocode} 
\usepackage{etoolbox}

\makeatletter
\newcommand*{\algline}[1][\algorithmicindent]{%
    \hspace*{.2em}%
    {\color{black!20}\vrule}%
    \hspace*{\dimexpr#1-.2em-.4pt}%
}

\renewcommand{\ALG@beginalgorithmic}{\offinterlineskip}

\newcount\ALG@printindent@tempcnta
\def\ALG@printindent{%
    \ifnum \theALG@nested > 0
        \ifx\ALG@text\ALG@x@notext 
        \else
            \unskip
            \ALG@printindent@tempcnta=1
            \loop
                \algline[\csname ALG@ind@\the\ALG@printindent@tempcnta\endcsname]%
                \advance \ALG@printindent@tempcnta 1
                \ifnum \ALG@printindent@tempcnta<\numexpr\theALG@nested+1\relax
            \repeat
        \fi
    \fi
}

\patchcmd{\ALG@doentity}{\noindent\hskip\ALG@tlm}{\ALG@printindent}{}{\errmessage{failed to patch}}
\makeatother

\algrenewcommand\algorithmicend{\strut\textbf{end}}
\algrenewcommand\algorithmicdo{\strut\textbf{do}}
\algrenewcommand\algorithmicwhile{\strut\textbf{while}}
\algrenewcommand\algorithmicfor{\strut\textbf{for}}
\algrenewcommand\algorithmicforall{\strut\textbf{for all}}
\algrenewcommand\algorithmicloop{\strut\textbf{loop}}
\algrenewcommand\algorithmicrepeat{\strut\textbf{repeat}}
\algrenewcommand\algorithmicuntil{\strut\textbf{until}}
\algrenewcommand\algorithmicprocedure{\strut\textbf{procedure}}
\algrenewcommand\algorithmicfunction{\strut\textbf{function}}
\algrenewcommand\algorithmicif{\strut\textbf{if}}
\algrenewcommand\algorithmicthen{\strut\textbf{then}}
\algrenewcommand\algorithmicelse{\strut\textbf{else}}

\algrenewcommand\algorithmicrequire{\strut\textbf{Input:}}
\algrenewcommand\algorithmicensure{\strut\textbf{Output:}}

\let\oldState\State
\renewcommand{\State}{\oldState\strut}

\newcommand{\scrF}{\mathcal{F}}

\newcommand{\scrFrep}{\scrF^{\text{rep}}}
\newcommand{\Frep}{F^{\text{rep}}}
\newcommand{\vrep}{\vc v^{\text{rep}}}
\newcommand{\verts}{\mathcal{V}}
\newcommand{\XX}{\mathcal{X}}
\newcommand{\YY}{\mathcal{Y}}
\newcommand{\neighbours}{\mathcal{M}}
\newcommand{\normals}{\mathcal{N}}
\newcommand{\defD}{\mathcal{D}}
\newcommand{\children}{\mathcal{C}}
\newcommand{\grp}{\mathcal{G}}
\newcommand{\grpP}{\mathcal{P}}
\newcommand{\setU}{\mathcal{U}}
\newcommand{\setT}{\mathcal{T}}
\newcommand{\setP}{\mathcal{P}}
\newcommand{\aopt}{a_{\text{opt}}}

\newcommand{\defeq}{\triangleq}
\newcommand{\anemptyset}{\varnothing}

\newcommand{\EmptyLine}{\Statex \vspace{2ex}}
\newcommand{\CommentFont}[1]{{\footnotesize\textcolor{gray!80!black}{#1}}}
\newcommand{\CommentLeader}{{\CommentFont{%
}}}
\newcommand{\CommentInlineText}[1]{{\CommentFont{%
    \CommentLeader\ignorespaces #1\unskip%
}}}
\newcommand{\CommentInline}[1]{\hfill \CommentInlineText{#1}}

\algrenewcommand{\alglinenumber}[1]{\footnotesize#1}

\makeatletter
\newcommand{\CommentLineEmpty}[2][0]{%
    \addtocounter{ALG@nested}{-#1}%
    \Statex\strut\ALG@printindent\relax\CommentInlineText{#2}%
    \addtocounter{ALG@nested}{#1}%
}
\makeatother

\newcommand{\proc}[1]{{\scshape\small #1}}
\newcommand{\class}[1]{{``#1''}}
\newcommand{\field}[1]{{``#1''}}
\newcommand{\lib}[1]{{\em #1}}

\newcommand{\eqnref}[1]{\eqref{#1}}

\newcommand{\R}{{\mathbb{R}}}
\newcommand{\Z}{{\mathbb{Z}}}
\newcommand{\vc}[1]{{\boldsymbol{#1}}}
\newcommand{\mat}[1]{{\boldsymbol{#1}}}

\newcommand{\AEnineM}{\text{AE}_9}
\newcommand{\AEnine}{$\AEnineM$\xspace}
\newcommand{\KtwelveM}{K_{12}}
\newcommand{\Ktwelve}{$\KtwelveM$\xspace}
\newcommand{\qq}{\sqrt3}

\DeclareMathOperator{\Conv}{Conv}
\DeclareMathOperator{\Vol}{Vol}
\DeclareMathOperator{\Aut}{Aut}
\DeclareMathOperator{\Stab}{Stab}
\DeclareMathOperator{\Perm}{Perm}
\DeclareMathOperator{\Diag}{Diag}
\DeclareMathOperator{\Gram}{Gram}

\newenvironment{alg}[1][tbp]{
    \begin{algorithm}[#1]
    \footnotesize
}{
    \end{algorithm}
}

\newenvironment{centeredtable}[2][tbp]{
    \begin{table}[#1]
    \footnotesize
    \caption{#2}
    \begin{center}
    
}{
    \end{center}
    \end{table}
}

\setlist[enumerate]{leftmargin=.5in}
\setlist[itemize]{leftmargin=.5in}

\headers{Quantizer Constants for Arbitrary Lattices}{D. Pook-Kolb, B. Allen, and E. Agrell}

\title{%
    Exact calculation of quantizer constants for arbitrary lattices%
    \thanks{%
        October 13, 2022.%
    }%
}

\author{%
    Daniel Pook-Kolb%
    \thanks{%
        Max Planck Institute for Gravitational Physics (Albert Einstein Institute),
        Callinstrasse 38, 30167 Hannover, Germany
        and
        Leibniz Universit\"at Hannover, 30167 Hannover, Germany
        (\email{daniel.pook.kolb@aei.mpg.de}, \email{bruce.allen@aei.mpg.de}).
    }
    \and
    Bruce Allen%
    \footnotemark[2]
    \and
    Erik Agrell%
    \thanks{%
        Chalmers University of Technology, Department of
        Electrical Engineering, SE-41296 Gothenburg, Sweden
        (\email{agrell@chalmers.se}).
    }%
}

\ifpdf
\hypersetup{
    pdftitle={Exact calculation of quantizer constants for arbitrary lattices},
    pdfauthor={D. Pook-Kolb, B. Allen, and E. Agrell}
}
\fi

\begin{document}

\maketitle

\begin{abstract}
    We present an algorithm for the computer-aided analytical construction
    of the Voronoi cells of lattices with known symmetry group.
    This algorithm is applied to the Coxeter--Todd lattice \Ktwelve as well as
    to a family of lattices obtained from laminating \Ktwelve.
    This way, we obtain a locally optimal lattice quantizer in $13$ dimensions
    representing a new best quantizer among the lattices with published exact
    quantizer constants.
\end{abstract}

\begin{keywords}
    lattice quantizer, quantizer constant, normalized second moment,
    Coxeter--Todd lattice,
    laminated lattice, Voronoi cell
\end{keywords}

\begin{MSCcodes}
    11H06, 11H56, 52B20, 52C07
\end{MSCcodes}

\section{Introduction}
\label{sec:intro}

Lattices are central to a whole spectrum of mathematical problems.
Among these are the sphere packing, kissing number, covering and quantization
problems, the latter being the focus in the present work.
A comprehensive review of these problems and descriptions of known classical
lattices are given by Conway and Sloane in \cite{splag}.

An $n$-dimensional lattice $\Lambda$ is a set of points in $\R^m$, $m \geq n$,
which is generated by integral combinations of $n$ linearly independent
vectors $\vc x_i \in \R^m$, where $i = 1, \ldots, n$.
We will adopt the convention that vectors $\vc x$ are treated as row vectors.
By a suitable rotation, it is always possible to embed $\Lambda$ in $\R^n$,
such that the matrix $\mat B$ of rows formed by the $n$ vectors $\vc x_i$
is an $n \times n$ invertible square matrix.
We can use $\mat B$ to express the lattice as
\begin{equation}\label{eq:lattice}
    \Lambda = \left\{ \vc z \mat B : \vc z \in \Z^n \right\} \;.
\end{equation}
$\mat B$ is called a {\em generator matrix} of the lattice $\Lambda$ and
$\vc x_1, \ldots, \vc x_n$ a set of {\em basis vectors}.
Non-overlapping translated copies of the parallelepiped defined by the
basis vectors cover all of $\R^n$, and their $n$-volume is defined as the
lattice's volume $\Vol(\Lambda)$.
For a square generator matrix, it is
$\Vol(\Lambda) = \lvert\det \mat B\rvert$.
Note that the volume is a characteristic of the lattice and is invariant under
rotations or a change of the set of basis vectors used to build up $\mat B$.

The main property of a lattice we are interested in is its
(dimensionless) {\em normalized second moment} or
{\em quantizer constant}
\begin{equation}\label{eq:G}
    G \defeq \frac{1}{n} \frac{E}{\Vol(\Lambda)^{2/n}} \,.
\end{equation}
Here, $E$ is the average squared distance of a uniform random distribution of
points in an $n$-dimensional ball of radius $r$
to the closest lattice point if $r$ is much larger than the lattice spacing.
The normalization by an inverse power of $\Vol(\Lambda)$ makes $G$ independent of the
overall scaling of $\Lambda$, whereas the factor $1/n$ makes lattices of
different dimensions have comparable values.
For example, the cubic lattice $\Z^n$ has $G = 1/12$ in any dimension $n$.
Minimizing $G$ for fixed $n$ is known as the {\em lattice quantizer problem}.

For a fixed dimension $n$,
finding the globally optimal lattice is a difficult task in general.
There are proofs of optimality only for lattices in dimensions up to three
\cite{gersho79, barnes83},
but it is believed that optimal quantizers have been found up to dimension $10$
\cite{agrell98,Agrell:2022jlo,Lyu2022better}.
An especially interesting case is the $9$-dimensional lattice \AEnine.
It was found approximately by Agrell and Eriksson in 1998 \cite{agrell98} using numerical
optimization, and was recently constructed exactly by Allen and Agrell
using analytic methods
\cite{allen21}.
This lattice, conjectured to be the optimal lattice quantizer in $9$
dimensions, is not one of the well-known classical lattices.
This is in striking contrast to the best lattice quantizers known in lower
dimensions, which are all classical.

A lower bound on $G$, known as Zador's lower bound
(or {\em sphere bound}, since it is found by calculating $G$ for an $n$-sphere),
is
\begin{equation}\label{eq:sphere-bound}
    G \geq \frac{\Gamma(1 + n/2)^{2/n}}{(n+2)\pi} \;,
\end{equation}
where $\Gamma$ is the gamma function.
Furthermore, there are conjectured lower (Conway and Sloane \cite{conway85})
and upper bounds (Zador \cite[Lemma~5]{zador82})
for the optimal lattice quantizer.
Zador's upper bound has been derived elegantly by
Torquato\footnote{%
    See \cite[\mbox{Footnote~[31]}]{allen2022performance} for a
    correction of the original argument.
}
\cite{torquato2010reformulation} as the quantizer constant of a
Poisson distribution of points in $\R^n$.

In many dimensions $n \geq 13$, no lattice was known that lies below the
Zador upper bound.
However, significant progress was made recently in \cite{Agrell:2022jlo},
where many new lattices are constructed with quantizer constants
representing new records in their respective dimension.
In particular, Zador's bound is now satisfied for lattices in dimensions $13$,
$14$, $17$ and $25$.

A key point made in \cite{Agrell:2022jlo} is that none of the newly
constructed lattices is optimal.
This presents a promising avenue for further exploration.

Additional improvements
were made in \cite{Lyu2022better} in dimensions $14$, $15$ and $18$--$23$,
where Zador's bound is now also satisfied in dimensions $15$, $18$ and $20$.

One of the major challenges is the exact calculation of $G$.
In the present work, we present our algorithm for solving this task and apply
it to the Coxeter--Todd lattice \Ktwelve
(see \cite[Section~4.9]{splag} and \cite{conway1983coxeter})
as well as to a $13$-dimensional lattice,
which is obtained by laminating \Ktwelve.

The first step towards transforming \eqnref{eq:G} into a solvable problem
is the following well-known observation.
Due to the translational symmetry of a lattice, $G$ can be calculated by
considering only a single lattice point $\vc x \in \Lambda$ and
the region
\begin{equation}\label{eq:Voronoi}
    \Omega(\vc x) \defeq \left\{
        \vc y \in \R^n :
            \|\vc y - \vc x\|^2 \leq \|\vc y - \vc x'\|^2,
            \ \ \forall \vc x' \in \Lambda
    \right\}
\end{equation}
of points closer to $\vc x$ than to any other lattice point.\footnote{%
    More precisely, $\Omega(\vc x)$ contains all points not further from
    $\vc x$ than from any other lattice point.
    However, we shall ignore sets of measure zero in this and the following
    statements.
}
For brevity, we will often use $\Omega$ to denote $\Omega(\vc 0)$.
This region is called the {\em Voronoi region} or {\em Voronoi cell} of
$\Lambda$.

The Voronoi cells of each lattice point are translated copies of each other,
centered around lattice points,
and they cover $\R^n$ without overlap,
just like the parallelepipeds formed by the basis vectors.
They hence have the same volume \cite[Prop.~2.2.1]{zamir14book}.

In fact, $\Omega$ is a convex polytope with facets lying halfway between the
origin and nearby lattice points.
The vectors connecting the origin with these points
are called {\em Voronoi-relevant vectors} or simply {\em relevant vectors}
and they lie perpendicular to the facets.
Let $\normals(\Omega)$ be the set of relevant
vectors. Then
\begin{equation}\label{eq:Voronoi-via-relevant}
    \Omega = \left\{
        \vc x \in \R^n :
            2\,\vc x \cdot \vc n \leq \|\vc n\|^2,\ \ \forall \vc n \in \normals(\Omega)
    \right\} \;.
\end{equation}

Using the Voronoi cell $\Omega$,
the average squared distance $E$ of a random point from any lattice point can
be written as the average squared distance of a point in $\Omega$ from the
origin, i.e.,
\begin{equation}\label{eq:E}
    E = \frac{1}{\Vol(\Lambda)} \int_\Omega \|\vc x\|^2\ d^n \vc x \;.
\end{equation}
$E$ is related to the (unnormalized) second moment $U$ of $\Omega$
about the origin via $E = U/\Vol(\Lambda)$.

A generalization of the second moment $U$
is the {\em covariance matrix} or {\em second moment tensor}
$\mat U$.
For a $d$-dimensional body $P_d$ in $\R^n$,
the second moment tensor about
a point $\vc x_0 \in \R^n$
is the $n \times n$ matrix
\begin{equation}\label{eq:Uab}
    \mat U(P_d, \vc x_0) \defeq \int_{P_d} (\vc x - \vc x_0)^T (\vc x - \vc x_0)\ d^d \vc x
    \;,
\end{equation}
where $d^d \vc x$
denotes a $d$-dimensional volume element.
The
trace of $\mat U \defeq \mat U(\Omega, \vc 0)$ is just the second moment $U$
of the Voronoi cell.

Zamir and Feder show in \cite{zamir96} that a lattice that minimizes $G$
has a second moment tensor $\mat U$ which is proportional to the identity matrix.
This result has been generalized
in \cite{Agrell:2022jlo}, where it is shown that it holds for any
{\em locally} optimal lattice $\Lambda$.
The geometric interpretation of the second moment tensor being
proportional to the identity matrix is that
the moment of inertia about any axis placed through the origin
is the same.

Throughout this work, we will denote
vectors $\vc x \in \R^n$ and matrices $\mat A \in \R^{k \times m}$
in
lower- and uppercase bold font, respectively.
Finite sets will be denoted by calligraphic uppercase letters such as
$\normals$, $\verts$ or $\scrF$, and their cardinality by, e.g., $|\scrF|$.
Definitions are introduced using ``$\defeq$''.

The rest of this paper is organized as follows.
In Sec.~\ref{sec:voronoi-construction}, we will describe an algorithm for
constructing the Voronoi cell of an arbitrary lattice with known
symmetry group.
Our presentation will revolve around the data structure we use for
representing the cell, which hopefully allows the reader to quickly get an
overview of the main ideas and our strategy.
After discussing the basic construction in four steps in a ``naive'' approach
in Sec.~\ref{sub:naive}, the remainder of Sec.~\ref{sec:voronoi-construction}
will detail the optimizations made possible by incorporating the symmetries.
The result is a hierarchy of faces, represented as objects in our algorithm.
Section~\ref{sec:second-moment} gives a brief account of the formulas that can
be used to calculate the second moment scalar and tensor of the cell.
In Sec.~\ref{sec:results}, we present the concept of product and laminated
lattices and finally apply our algorithm to the Coxeter--Todd lattice
\Ktwelve and a lamination of \Ktwelve.
We conclude in Sec.~\ref{sec:conclusions} with a summary and
discussion of our method and the results.

\section{Constructing the Voronoi cell}
\label{sec:voronoi-construction}

\begin{figure*}\centering
    \includegraphics[width=\linewidth]{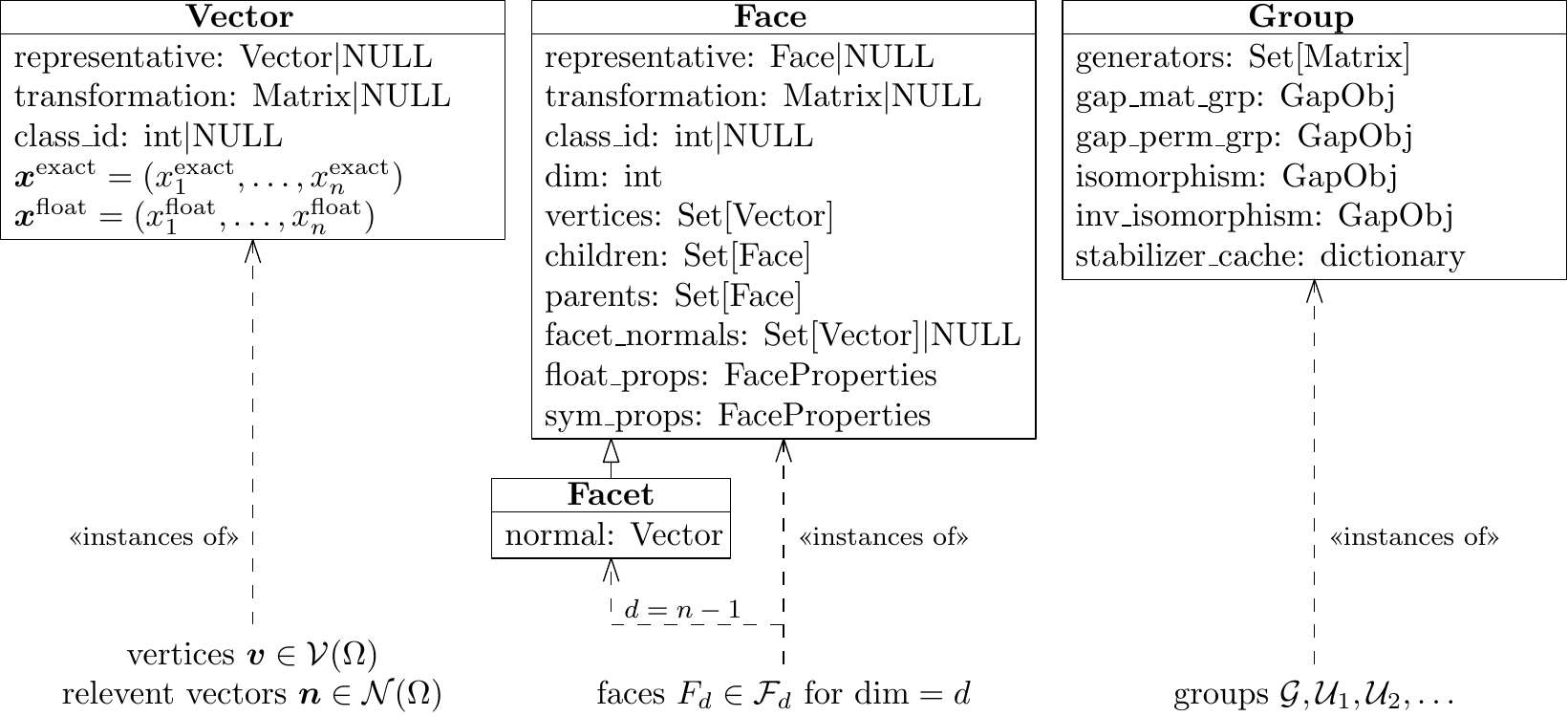}%
    \caption[]{\label{fig:data-structure}%
        Data types we use to construct and represent the face hierarchy of
        a Voronoi cell.
        The \field{class\_id} fields are populated only for one representative per
        equivalence class, while \class{representative} and
        \class{transformation} only hold data for the remaining elements.
        The \class{Group} class is a thin wrapper around \lib{GAP}
        group objects and related operations.
        The two
        fields
        \field{float\_props} and \field{sym\_props} as well as
        the class \class{FaceProperties}
        are discussed in Sec.~\ref{sec:second-moment}.
        See the main text for details.
    }
\end{figure*}

In this section, we present our algorithm for constructing the Voronoi
cell of an $n$-dimensional lattice $\Lambda$.
The goal is to build a data structure that allows the exact calculation of its
second moment (scalar and tensor).
The method presented here is derived from the one used in \cite{allen21}
but makes greater use of the symmetries of the lattice.

The basic idea is as follows.
From \eqnref{eq:Voronoi-via-relevant} it is clear that points on the
$(n-1)$-dimensional faces (the {\em facets}) of $\Omega$ saturate the inequality
for at least one relevant vector $\vc n \in \normals(\Omega)$.
The facets are themselves convex polytopes, with $(n-2)$-dimensional subfaces
lying at the intersections of facets.
By recursively intersecting subfaces, we obtain faces of ever lower dimension
until we arrive at the $0$-faces, the {\em vertices}.
We shall denote the set of vertices of an arbitrary polytope $P_d$ of
dimension $d$ as $\verts(P_d)$.

Above and in the following, we use the term {\em subface} to refer to
a $d$-face $F_d$ that is a proper subset of a $d'$-face $F_{d'}$,
where $d' > d$.
A {\em child} or {\em child face} is a subface $F_d$ of a face $F_{d'}$ of
exactly one lower dimension, $d' = d+1$.
Similarly, a {\em parent} or {\em parent face} of a face $F_d$ is a face
$F_{d'}$ such that $F_d$ is a child of $F_{d'}$.

The building blocks of our data structure are shown in
Fig.~\ref{fig:data-structure}.
In the end, we will have a hierarchy of \class{Face} objects, related to each
other via parent--child relationships.
The recursive formulas we introduce in Section~\ref{sec:second-moment} for
calculating the scalar and tensor second moment can then easily be applied to this
data structure.

In what follows,
we will often switch between mathematical statements
about, e.g., equivalence classes, stabilizers or convex polytopes
and the representations of these concepts via objects in the data structure.
Whenever possible, we use the same notation for mathematical objects and those
in our code.
For example, a $d$-face $F_d$ is a $d$-dimensional convex polytope,
i.e., $F_d \subset \R^n$, where ``$\subset$'' denotes a proper subset.
But it may also refer to an object of type \class{Face} that stores a set of,
say, $N$ vertices $\{\vc v_1, \ldots, \vc v_N\}$ as \class{Vector} objects,
which determine the face via their convex hull,
$F_d = \Conv(\{\vc v_1, \ldots, \vc v_N\})$.
The distinction between these two cases is made explicit only when it is not
clear from context.
As mentioned above, (nested) sets or lists of faces will often be written as
calligraphic $\scrF$, while individual objects are always set in a
non-calligraphic font.
Within the pseudocode of listed algorithms, spelled-out variables names such
as ``{\em faces}'' are sometimes used for clarity.

The three classes shown in Fig.~\ref{fig:data-structure} form the basis of our
data structure.
We defer the discussion of the full set of shown fields to the
points where they become relevant.
For now, we will focus on their general roles.

\class{Vector}:
These objects represent points in $\R^n$ and vectors between points.
The components are stored as exact expressions and in floating point
representation.
The latter is used for quick preliminary and verification calculations
as well as for certain decision processes.\footnote{
    An example of such a decision is determining if a set of vectors is
    linearly independent.
    Other cases will be discussed in the following subsections.
}
Relevant vectors and vertices that define faces are stored as \class{Vector}
objects.

\class{Face}:
Instances of this class represent the various faces of the Voronoi cell.
They are defined by a set of vertices stored as \class{Vector} objects.
For each face, its parent faces and child faces are stored, to enable
navigation in the hierarchy of faces.
Note that \class{Face} has a subclass \class{Facet} that adds storage for the
relevant vector, its \field{normal}, which defines the plane it lies in
(see below).

Note that the vertices appear in Fig.~\ref{fig:data-structure} both
as \class{Vector} and as \class{Face} objects, since
the $0$-face objects $F_0$ store just a single vector
$\vc v \in \verts(\Omega)$ in their \field{vertices} field.
This reflects the two roles played by vertices in our data structure:
Certain sets of vertices define faces, while each vertex is also a
hierarchical entity with parent $1$-faces.

\class{Group}:
Symmetry groups and certain subgroups will be stored as objects of this type.
It is implemented as a wrapper for groups in \lib{GAP}\footnote{%
    Our use of \lib{GAP} requires a more recent version than
    4.11.1.
    Specifically, we use the commit
    \cite{GAPRepo}.
}
\cite{GAPSoftware},
which provides a programming language for computational group theory.
The various fields will be discussed further below.

The notion of symmetries and equivalence of vectors and faces will be captured
by the fields
\field{representative}, \field{transformation} and \field{class\_id}, which
are described in Sec.~\ref{sub:symmetries-general}.

The majority of our code is written in the \lib{Python} programming language
and the interface to \lib{GAP} is provided by the Python module
\lib{gappy} \cite{gappySoftware}.
Note that the types \class{Set}, used in Fig.~\ref{fig:data-structure}, as
well as \class{List} and \class{Dictionary}, which we use in the algorithms
below, are precisely the {\em set}, {\em list} and {\em dict} types used in
Python.
For example, a \class{Set} object is an unordered collection of items with
efficient ways to compute intersections or to test for membership.
Similarly, a \class{Dictionary} is an object that maps a set of {\em keys} to
corresponding {\em values},
akin to lists mapping non-negative integers (the indices) to values.
In contrast to lists, the keys of a \class{Dictionary} can be, e.g., strings or
even objects of type \class{Vector} or \class{Face}.
In Fig.~\ref{fig:data-structure}, the type of objects stored in a \class{Set}
is given in square brackets.

It is common in Python that assigning an object to a variable effectively
creates a pointer to the underlying object.
This mechanic is implied in all algorithms shown here.
One consequence of this is that, e.g., the set of vertices each \class{Face}
object stores consists of \class{Vector} objects from the common pool of all
vertices in $\verts(\Omega)$.
That is, no copies are created and stored.

We now start without considering symmetries and present a conceptually
simple method of constructing the face hierarchy of the Voronoi cell.
This will introduce the basic tasks and highlight the most expensive steps.
Our algorithm then builds upon this method and introduces efficient alternatives.

\subsection{The naive approach}
\label{sub:naive}

An explicit construction of the Voronoi cell may be carried out using the
following four steps, where steps 1, 3 and 4 are very similar to those
presented in \cite{allen21}.

{\em Step 1.}
One first finds all relevant vectors $\vc n_i \in \normals(\Omega)$.
An efficient algorithm,
which is based on \cite[Chapter~21, Theorem~10]{splag},
is described in \cite{AEVZ}.
The relevant vectors lie orthogonal to the facets, which in turn lie in the
hyperplanes
\begin{equation}\label{eq:facet-planes}
    E_i \defeq \left\{ \vc x \in \R^n : 2\,\vc x \cdot \vc n_i = \|\vc n_i\|^2 \right\}
    \;.
\end{equation}
It was shown \cite{minkowski1989allgemeine,Voronoi4}
that there are at
most $2(2^n-1)$ relevant vectors.

Without considering any symmetries for now, we store each of the vectors
$\vc n_i$ as \class{Vector} object with no data for the fields
\field{representative}, \field{transformation} and \field{class\_id}.
On the other hand, the facets are stored as objects of type \class{Facet}.
At this point, they contain data in their \field{normal} and \field{dim}
fields, the latter being set to $n-1$, which is the dimension of the facets of
an $n$-dimensional polytope.

{\em Step 2.}
The next step is to find all vertices $\verts(\Omega)$ of the Voronoi cell.
It is easy to see that the vertices of $\Omega$ are {\em holes} of $\Lambda$,
i.e., points that have a locally maximal distance to the nearest lattice
points.
In principle, they can be obtained by taking all possible combinations of $n$
linearly independent\footnote{
    By \eqref{eq:facet-planes}, $n$ planes $E_i$ intersect in the points
    $\{\vc x \in \R^n : 2\vc x \mat M^T = \Diag(\mat M\mat M^T)\}$,
    where the $n \times n$ matrix $\mat M$ consists of rows formed by the $n$
    relevant vectors $\vc n_i$ corresponding to the planes $E_i$.
    These planes intersect in exactly one point if and only if $\mat M$ has full rank,
    i.e., if the relevant vectors are linearly independent.
}
vectors $\vc n_i \in \normals(\Omega)$ and intersecting the corresponding planes $E_i$.
A point $\vc x$ obtained in this way is a vertex if and only if
it is an element of the Voronoi cell, i.e., if
\begin{equation}\label{eq:vertex-conditions}
    2\,\vc x \cdot \vc n_i \leq \|\vc n_i\|^2,\ \ \forall \vc n_i \in \normals(\Omega)\;.
\end{equation}
As in step 1, the vertices are stored as \class{Vector} objects without any
symmetry-related data.

{\em Step 3.}
We next assign each facet its set of vertices by evaluating
\eqref{eq:vertex-conditions} for fixed $i$ and all $\vc x \in \verts(\Omega)$,
and checking for equality.
This check is done using floating point calculations
as
\begin{equation}\label{eq:vertex-in-facet-float}
    \vc x \in \verts(F_{n-1}^i)
    \quad\Longleftrightarrow\quad
    \big|2 \vc x \cdot \vc n_i - \|\vc n_i\|^2\big| \leq \varepsilon \;,
\end{equation}
where $F_{n-1}^i$ is the facet associated with $\vc n_i$ and $\varepsilon$ a
small tolerance to account for floating point roundoff errors.
The resulting sets of vertices then populate the \field{vertices} field of the
corresponding facet objects.

\begin{figure}\centering
    \includegraphics{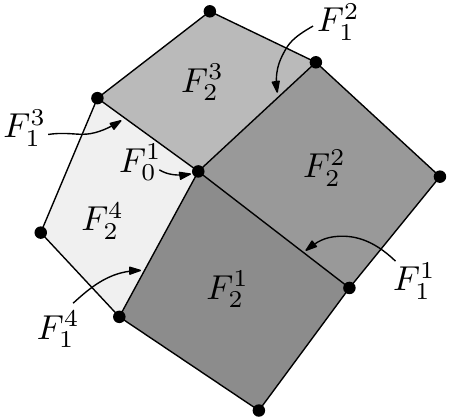}%
    \caption[]{\label{fig:intersection-too-low-dim}%
        Four $2$-faces meeting at a common vertex.
        The intersection of the $2$-faces $F_2^1$ and $F_2^3$
        is the $0$-face $F_0^1$, which is a subface but not a child
        face of either.
        Conversely, $F_2^1$ and $F_2^4$ intersect at a child face $F_1^4$.
    }
\end{figure}

{\em Step 4.}
The $(n-2)$-dimensional faces of the Voronoi cell are
obtained by intersecting the sets of vertices of pairs of facets.
This is easy to do with the data structure in Fig.~\ref{fig:data-structure}.
We intersect the \field{vertices} fields of the two facet objects and
assign the result to the \field{vertices} field of a new face object.

In many cases, however, this is not guaranteed to produce a child face.
For example, the intersection will be empty if the two facets do not touch.
Similarly, their intersection may be a subface that is not a (direct) child.
Such a case is visualized in Fig.~\ref{fig:intersection-too-low-dim} where two
$2$-dimensional facets meet in a $0$-face.
We therefore numerically evaluate the dimension of the convex hull of the
new smaller set of vertices.
This is done by constructing all vectors connecting one of these vertices with
the others and counting the number of linearly independent vectors.
Again, we can use the floating point components of the \class{Vector} objects
for this otherwise expensive calculation.
If the dimension is $n-2$, the hull is a child face and we populate its
\field{parents} field and update
the respective facets' \field{children} fields.
If the dimension is less than $n-2$, the face is discarded.

It is possible that distinct combinations of faces intersect in the same child
face.
For example, the $0$-face $F_0^1$ in Fig.~\ref{fig:intersection-too-low-dim}
is the intersection of any pair of the $1$-faces
$F_1^1, \ldots, F_1^4$.
We check for that case by comparing the sets of vertices of any newly found
child face $F_{d-1}$ against previously found ones.
If there is a match $F_{d-1}'$, the \field{parents} field of $F_{d-1}'$ and
the \field{children} field of the intersected faces are updated, respectively,
and the duplicate $F_{d-1}$ is discarded.

Recursively intersecting child faces as above produces the hierarchy of all
lower-di\-men\-sio\-nal faces.

There are two main problems of this naive approach.
First, steps 2 and 4 involve explicit iteration over all
possible combinations of quantities which may rapidly grow in number with
the dimension $n$ of the lattice.
%
As an example, \AEnine has $370$ relevant vectors and so in step 2 one would
need to check more than $3 \times 10^{17}$ combinations of $n=9$ vectors
$\vc n_i \in \normals(\Omega)$ for linear independence and then intersect the
planes $E_i$.
Similarly, step 4 would involve computing and checking about
$7 \times 10^{12}$ intersections of two faces to construct the full hierarchy
of $7\,836\,067$ faces.
%
This gets much worse in higher dimensions.
\Ktwelve, for example, has $4\,788$ relevant vectors and hence would require
checking almost $3 \times 10^{35}$ combinations in step 2.

The other problem is that the total number of faces may become too large to
feasibly perform any recursive calculations involving all those faces.
Even though it is still possible to recursively calculate the second moment for the nearly
$8$ million faces of \AEnine, this becomes exceedingly difficult in higher
dimensions.

In the next sections, we will therefore show how this naive method can be
improved significantly.
The main theme of these improvements will be to use symmetries of the
Voronoi cell to greatly reduce the number of objects we need to
explicitly track.

\subsection{Symmetries of a lattice}
\label{sub:symmetries-general}

When viewed as a discrete subset of $\R^n$, we can ask which rotations
$O(n)$ take the lattice $\Lambda$ into itself.
The subset of $O(n)$ that does this is called the
{\em symmetry group} or {\em automorphism group} $\Aut(\Lambda)$
of the lattice, and we write it as
\begin{equation}\label{eq:aut}
    \grp \defeq \Aut(\Lambda) \defeq \{ g \in O(n) : g \Lambda = \Lambda \} \;.
\end{equation}
The symmetry group of a lattice is always finite
(see, e.g., \cite[Section~3.4.1]{splag} and \cite[Theorem~1.4.2]{martinet13book}).

The (left) action of a symmetry $g \in \grp$ on a row vector $\vc x \in \R^n$
will be written as
\begin{equation}\label{eq:group-acting-on-vector}
    g \vc x \defeq \vc x \mat M_g \;,
\end{equation}
where $\mat M_g^T$ is the $n \times n$ matrix representing $g$ on $\R^n$,
i.e., $g$ acts via $\mat M_g^T$ on column vectors on the left.\footnote{
    We use the libraries
    \lib{NumPy} \cite{van_der_Walt_NumPy}
    and
    \lib{GAP} \cite{GAPSoftware},
    which work with left actions and right actions, respectively.
    In our case, it is trivial to convert between the conventions by
    simply transposing the matrices.
    Hence, the transpose of the matrix
    representing $g$ is multiplied from the right in
    \eqref{eq:group-acting-on-vector} to define the left action here.
}

The action of $\grp$ on faces and any other set of points in $\R^n$ is defined
analogously by applying \eqref{eq:group-acting-on-vector} to each point.
For a convex $d$-dimensional polytope $P_d$ with vertices $\verts(P_d)$, this
immediately implies that
\begin{equation}\label{eq:g-on-polytope}
    g P_d = \Conv(g\verts(P_d)) \;.
\end{equation}

The {\em orbit} of a vector $\vc x$ under $\grp$ is the set
\begin{equation}\label{eq:orbit-of-vector}
    \grp \vc x \defeq \{g\vc x : g \in \grp\} \;.
\end{equation}
We say that two vectors $\vc x$ and $\vc x'$ are {\em equivalent under $\grp$}
if and only if they lie in the same orbit,
i.e., if there exists a $g \in \grp$ such that $\vc x' = g \vc x$.
In that case we will write $\vc x \sim \vc x'$.
In the present work, all equivalence relations will originate from group
actions, and so we will use the terms ``orbit'' and ``(equivalence) class''
interchangeably.

The notion of orbits and equivalence generalizes in the obvious way to any set
acted on by $\grp$, in particular to the faces of the Voronoi cell.

To see that a face always remains within $\Omega$ under $\grp$,
consider a symmetry $g \in \grp$.
Clearly, since by definition $g\Lambda = \Lambda$, the set of all Voronoi
cells remains invariant as well.
In fact, if $\vc x \in \Lambda$ is a lattice point, then
$g\Omega(\vc x) = \Omega(g\vc x)$.
However, since the origin is fixed by $g$, we have
$g\Omega(\vc 0) = \Omega(\vc 0)$ and so $g$ is a symmetry of
$\Omega$.\footnote{
    The converse is also true:
    Any $g \in O(n)$ taking $\Omega$ into itself must be in $\grp$.
    This follows from the fact that translated copies of $\Omega$
    form a tessellation of $\R^n$.
    Since $\Lambda$ consists of the union of center points of the cells, and
    since $g$ maps $\Omega$ to $\Omega$, the tessellation is invariant and so
    we have $g\Lambda = \Lambda \implies g \in \grp$.
}
This implies that the complete sets of vertices, relevant vectors and faces are
individually invariant under $\grp$.

In our data structure, the symmetry group $\grp$ is represented by
the \class{Group} class shown in Fig.~\ref{fig:data-structure}.
We store a set of matrices that generate the full group in the
\field{generators} field.
Since any group element $g \in \grp$ can be written as a finite product of
such generators, this holds the complete information necessary to construct the
group.
In practice, this set can be relatively small even for large groups.
For example, the Coxeter--Todd lattice \Ktwelve has a symmetry group of order
$78\,382\,080$ \cite[p.~129]{splag},
which we generate with just three matrices
(see Sec.~\ref{sub:K12}).

The generator matrices are used to construct a \lib{GAP} group object using
the already mentioned \lib{GAP} system via \lib{gappy}.
This object is stored in the \field{gap\_mat\_grp} field of our \class{Group}
class.
Since some of the operations in \lib{GAP} are implemented much more
efficiently for permutation groups than for matrix groups,
we also let \lib{GAP} construct a permutation group $\grpP$ that is
isomorphic to $\grp$. This is stored in the
\field{gap\_perm\_grp} field alongside the isomorphism and its inverse.

In our method, only a small number of instances of the \class{Group} class
will be created.
In particular, the full symmetry group $\grp$ is such an instance.
We furthermore need several subgroups $\setU_i$ of $\grp$, which are discussed
in Sec.~\ref{sub:faces}.
These are stored as \class{Group} instances as well.
However, elements $g$ of the full group or of a subgroup will just be
stored as matrices.

The equivalence of vertices, relevant vectors and faces is represented
as follows.
Each vector and face has fields
\field{representative}, \field{transformation} and \field{class\_id}.
For any equivalence class, only one arbitrary element
(usually the first one encountered) is chosen as the canonical representative
of its class.
This element will get a unique integer as \field{class\_id} and no data for
the other two fields.
Any equivalent item will then store a pointer to this chosen representative
and the group element (in matrix form) taking the representative into the item.
It will have data in the
\field{representative} and \field{transformation} fields and no data
in \field{class\_id}.

In a concrete implementation, one may choose to have two subclasses of,
e.g., the class \class{Vector}, one for the representative and one for the
transformed equivalent vectors.
This would avoid reserving redundant storage for the unused fields.
For the purposes of discussing our algorithm, however, we will assume that
the classification information is stored as shown in
Fig.~\ref{fig:data-structure}.

\subsection{Finding the vertices}
\label{sub:vertices}

Knowledge of a lattice's symmetry group $\grp$ provides an
immediate optimization of step 2 of the naive strategy in Sec.~\ref{sub:naive}.
The idea is that once a single vertex $\vc v$ has been found,
the orbit of $\vc v$ under $\grp$ provides a large number of additional
vertices.

Furthermore, instead of forming intersections of facets, it is more efficient
to numerically search
within $\Omega$
for local maxima of $\|\vc x\|^2$.
This can be done using linear programming, with $\Omega$ characterized by
\eqref{eq:vertex-conditions}.
As in \cite[Section~2]{allen21}, we solve
$\max_{\vc x \in \Omega}(\vc c \cdot \vc x)$
multiple times, but rather than choosing $\vc c$ uniformly, we find it more
efficient to set it to a relevant vector $\vc n \in \normals(\Omega)$ with a
small random perturbation.

To obtain an exact expression for a vertex $\vc v$ found numerically,
we first use \eqref{eq:vertex-in-facet-float} to collect all relevant vectors
$\vc n_i \in \normals(\Omega)$ of the facets it lies in.
Since a vertex lies in at least $n$ facets, this set will contain at least $n$
vectors.
If it is larger, we select $n$ linearly independent ones.
The intersection of the corresponding planes $E_i$ is formed as in step 2,
yielding the exact expression of the vertex $\vc v$.

Further vertices are obtained by utilizing the translational symmetries of
$\Lambda$ as follows:
Let $\vc v$ be a vertex and let $\neighbours(\vc v)$ be the set of nearest
lattice points of $\vc v$.
These can be found using the algorithms presented in \cite{AEVZ}.
Since $\vc v$ is a vertex of the Voronoi cell, the origin is one of those lattice
points.
More generally, $\vc v$ is a vertex of all the Voronoi cells
$\Omega(\vc n)$ for any
$\vc n \in \neighbours(\vc v)$.
However, since all the Voronoi cells are congruent and just translated copies of
each other, we have
\begin{equation}\label{eq:translated-holes}
    \vc v \in \verts(\Omega) \implies \vc v - \vc n \in \verts(\Omega),
    \ \ \forall \vc n \in \neighbours(\vc v)
    \;.
\end{equation}
Vertices found via \eqref{eq:translated-holes} need not be equivalent
under $\grp$, as we have used the translational symmetries here.
These are, by definition, not included in the automorphism group $\grp$.
Any new vertex thus leads to further vertices by constructing its
orbit under $\grp$.

Whenever we construct the orbit of a vertex, we first define that vertex as
the fixed representative of its class and assign it a new \field{class\_id}.
Upon applying the group to this representative, we store the representative
and the group element in the respective fields of our \class{Vector} class
(see Fig.~\ref{fig:data-structure}).
We use the sets of vertices in the orbits found thus far
in order to quickly check if any newly found vertex represents a new or a
known class.
Vertices in known classes are then discarded.

The numerical search for vertices is continued with random initial conditions
until no new classes of vertices appear.
We do not have a strict criterion for stopping this search.
However, when many iterations produce only known classes, one may proceed with
the analysis under the assumption that all vertices have been found.
At a later point, a side product of our calculations is the volume of the
vertices' convex hull.
Comparing this volume with the known volume $\Vol(\Lambda) = \lvert\det \mat B\rvert$
provides an unambiguous consistency check, since
the convex hull of every proper subset of the full set
of vertices also has a strictly smaller volume than the full Voronoi cell.

This makes the numerical task of determining all vertices feasible even for
higher dimensions.
For example, the symmetry group of laminated \Ktwelve
(see Sec.~\ref{sub:laminatedK12})
partitions the
$52\,351\,632$
vertices of the Voronoi cell into $482$ different classes.
By using \eqref{eq:translated-holes}, one can find vertices of each of these
classes from just $74$ numerically found vertices.
This means that a numerical random search needs to find at least one of each
of the $74$ subsets of vertices, which can be accomplished on a single core
within less than a day.

\subsection{Constructing the hierarchy of faces}
\label{sub:faces}

The most challenging of the steps to optimize is
the construction of the hierarchy of faces,
i.e., step 4 of the naive approach in Sec.~\ref{sub:naive}.
We will describe our solution to this problem in two parts.
The first part is discussed in this section
and focuses on optimizing the structure of the hierarchy,
given knowledge about the equivalence of faces.
This is followed in Sec.~\ref{sub:face-equivalence} by an algorithm
for evaluating that equivalence.

\subsubsection{Strategy for constructing the face hierarchy}
\label{sub:faces:overview}

\begin{figure}\centering
    \includegraphics{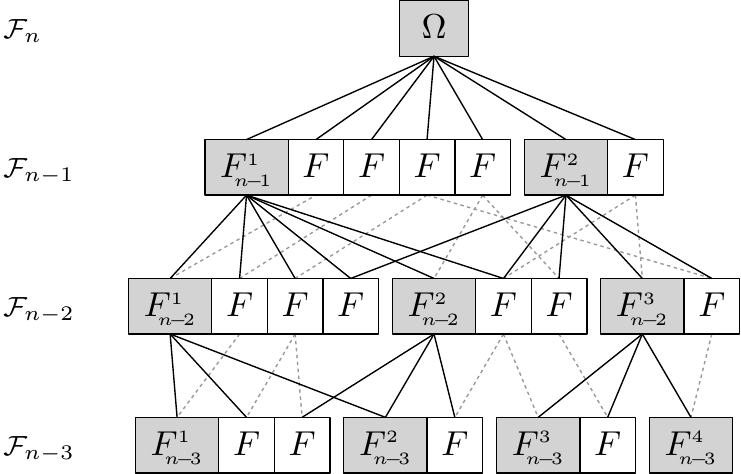}%
    \caption[]{\label{fig:hierarchy}%
        Faces explicitly constructed in
        Alg.~\ref{alg:ConstructFaces}.
        The rows of touching boxes indicate equivalent faces, each containing one
        arbitrary fixed representative on a gray background.
        Solid lines show the parent--child relationships between these
        representatives and their complete sets of children,
        while dashed lines indicate
        the remaining parent--child relationships.
    }
\end{figure}

Instead of building the full hierarchy of all faces, we will show that for the
calculation of the second moment tensor \eqref{eq:Uab} it is
sufficient to explicitly consider only a small subset of faces,
exploiting the symmetries.

The basic construction is visualized in Fig.~\ref{fig:hierarchy}.
It shows the faces of various dimensions $d$ as boxes in
different rows.
Within each row, faces that are equivalent under $\grp$ lie adjacent
to each other, while small gaps demarcate inequivalent subsets.
For example, the figure shows nine $(n-2)$-faces partitioned into three classes.
In each class, we arbitrarily select one face as representative,
shown as a box with gray background.
The children of these representative faces are indicated by solid lines.
In fact, except for the top layer $\scrF_n$, the sets $\scrF_d$ of $d$-faces
that we construct consist of {\em only} those children; no other face is
needed.

We point out a few important properties of this structure:
\begin{enumerate}[label=(\emph{\roman*})]
    \item\label{enum:all-children}
        The sets of children of each chosen representative face is
        {\em complete}, i.e., the face hierarchy contains all their child faces
        (solid lines in Fig.~\ref{fig:hierarchy}).
    \item\label{enum:all-classes}
        {\em At least one} face of each class of
        faces is constructed.
    \item
        At least one parent is a representative.
        However, in general we do {\em not} have a full set of parents.
    \item\label{enum:few-children}
        In general, only few of the children of faces that are not chosen as
        representative are constructed
        (dashed lines in Fig.~\ref{fig:hierarchy}),
        even though they clearly have the same number of children as all others
        in their class.
        There may even be cases, like the last face in the class of $F^1_{n-2}$, where
        none of its children are constructed.
\end{enumerate}
Points \ref{enum:all-children} and \ref{enum:all-classes} are essential for
the correctness of our results.
After describing our algorithm in detail, we will therefore give proofs that
the resulting structure possesses these properties.

One might be worried that the arbitrary choice of a representative in each
class of faces may lead to intersections being missed.
For example, consider two inequivalent representatives $F$ and $F'$
and their orbits under $\grp$.
If $g_1 F$ intersects with $g_2 F'$ in a child face
$F_C$, i.e.,
\begin{equation}\label{eq:F_C-no-transform}
    F_C = g_1 F \cap g_2 F' \;,
\end{equation}
for some $g_1, g_2 \in \grp$,
then how do we know that the children of $F$ and $F'$ contain a face
equivalent to $F_C$?
Properties \ref{enum:all-children} and \ref{enum:all-classes} precisely
say that this is the case.
In this particular situation, we can simply transform $F_C$
with $g_1^{-1}$ to see that
\begin{equation}\label{eq:F_C-transformed}
    g_1^{-1} F_C \subset F
\end{equation}
and so $F_C$ is equivalent to $g_1^{-1}F_C$, which is a child of $F$.

The above construction greatly reduces the total number of faces to consider
compared to the full set of faces.
For example, the $9$-dimensional lattice \AEnine has
$7\,836\,067$ faces in total which fall into $170$ classes.
By including only the child faces of the representatives, our method results
in about $2\,000$ faces to construct.\footnote{
    The precise number depends on which faces are chosen as fixed
    representatives of their class, since multiple representatives may share
    common child faces.
}

We remark that we lose easy access to the information about the total number
of $d$-faces that the Voronoi cell has.
We suspect that a combinatorial argument may allow for restoring this
number.
This may also involve information about the subfaces that occur in
intersections but are discarded due to having too low dimension
(see step 4 in Sec.~\ref{sub:naive}).
However, we have not been able to resolve this question, which is not critical
for the computation of quantizer constants.

\subsubsection{The basic algorithms}
\label{sub:faces:algorithms}

\begin{figure}\centering
    \includegraphics{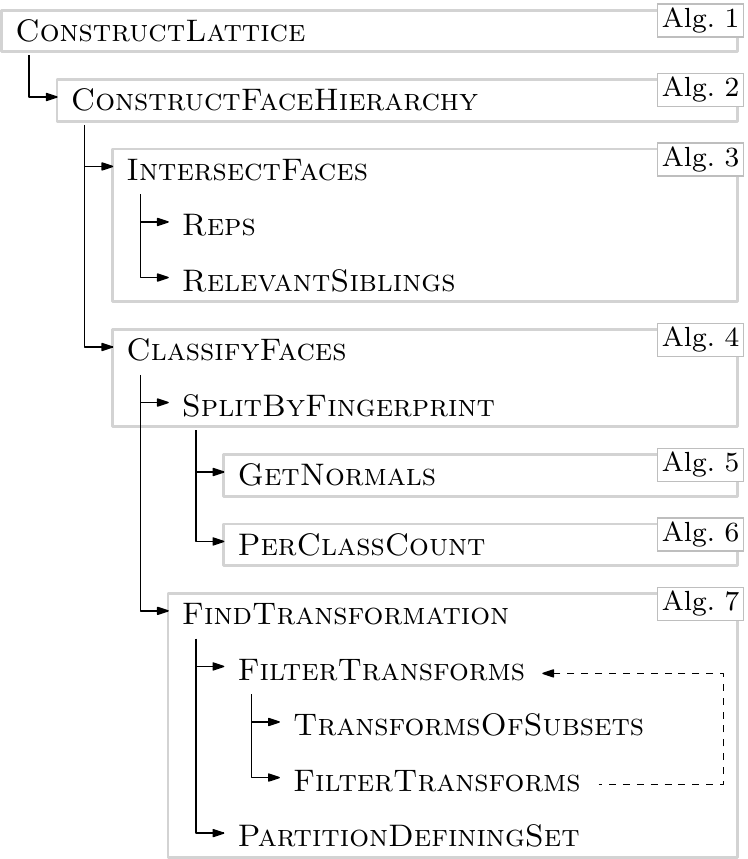}%
    \caption[]{\label{fig:alg-map}%
        Call graph showing the relationships between the procedures we
        present in this paper.
        Arrows point from the calling to the called procedure.
        The labeled boxes indicate in which of our algorithm listings the
        procedures can be found.
    }
\end{figure}

We will now introduce our algorithms that construct the hierarchy of faces.
They are split into several procedures, which are shown in this section and
Sec.~\ref{sub:face-equivalence}.
Fig.~\ref{fig:alg-map} shows how the various procedures are related and in
which algorithm they can be found.

\begin{alg}
    \caption{\raggedright
        Setup routine for building the face hierarchy.
    }\label{alg:Setup}
    \begin{algorithmic}[1]
        \Procedure{ConstructLattice}{\mbox{}}
            \State $n$ $\gets$ dimension of the lattice
                \label{ConstructLattice:n}
            \State {\em gens} $\gets$ Set of generator matrices of $\grp$
                \label{ConstructLattice:gens}
            \State $\normals$ $\gets$ Set of relevant vectors (step 1 of Sec.~\ref{sub:naive})
            \State $\verts$ $\gets$ Set of vertices (Sec.~\ref{sub:vertices})
            \State \Return \Call{ConstructFaceHierarchy}{$n$, {\em gens}, $\normals$, $\verts$}
        \EndProcedure
    \end{algorithmic}
\end{alg}

Several preparatory steps are required before the face hierarchy can be built.
For definiteness, they are shown in Alg.~\ref{alg:Setup} and build on the
results of the previous steps 1 and 2 of Sec.~\ref{sub:naive}
(with step 2 being replaced by Sec.~\ref{sub:vertices}).
Alg.~\ref{alg:Setup} is specific to each lattice we
analyze.
For the Coxeter--Todd lattice \Ktwelve, $n$ is set to $12$ in
line~\ref{ConstructLattice:n} and the generator matrices in
line~\ref{ConstructLattice:gens} are those given below in
\eqnref{eq:K12-group-gens}.
With the preparations done, the final step is calling the generic procedure
\proc{Construct\-Face\-Hierarchy}, which is shown in
Alg.~\ref{alg:ConstructFaces} and implements our algorithm for constructing
the hierarchy of faces.

\begin{alg}
    \caption{\raggedright
        Construct a hierarchy of faces.
        See the main text for details.
    }\label{alg:ConstructFaces}
    \begin{algorithmic}[1]
        \Procedure{ConstructFaceHierarchy}{$n$, {\em gens}, $\normals$, $\verts$}
            \State $\grp \gets$ new GapGroup
                \CommentInline{Create symmetry group object}
            \State $\grp$.generators $\gets$ {\em gens}
            \State {\em facets} $\gets$ facets using $\normals$ and $\verts$ as in step 3 of Sec.~\ref{sub:naive}
                \label{ConstructFaceHierarchy:facets}
            \State $F_n \gets$ new Face
                \label{ConstructFaceHierarchy:newFace}
                \CommentInline{Create $n$-face and populate its fields}
            \State $F_n$.dim $\gets$ $n$
            \State $F_n$.class\_id $\gets$ $0$
            \State $F_n$.vertices $\gets \verts$
            \State $F_n$.children $\gets$ {\em facets}
                \label{ConstructFaceHierarchy:newFaceDone}
            \State $\scrF \gets [\{\}, \ldots, \{\}, F_n\text{.children}, \{F_n\}]$
                \label{ConstructFaceHierarchy:initResultList}
                \CommentInline{$\scrF$ is a list of sets $\scrF_0, \scrF_1, \ldots, \scrF_n$}
            \For{$d \gets n-2$ to $0$} \label{ConstructFaceHierarchy:loop}
                \CommentInline{Iterate from $n-2$ to $0$-faces}
                \State \Call{ClassifyFaces}{$\scrF_{d+1}$, $\grp$, {\em facets}}
                    \label{ConstructFaceHierarchy:classify}
                    \CommentInline{Classify the parent faces $\scrF_{d+1}$}
                \State {\em children} $\gets$ \Call{IntersectFaces}{$\scrF_{d+1}$}
                    \label{ConstructFaceHierarchy:intersect}
                    \CommentInline{Create the children by intersecting parents}
                \State $\scrF_{d}$ $\gets$ {\em children}
                    \label{ConstructFaceHierarchy:assign-children}
                    \CommentInline{Store $d$-faces in result list}
            \EndFor
            \State \Return $\scrF$
        \EndProcedure
    \end{algorithmic}
\end{alg}

Alg.~\ref{alg:ConstructFaces} itself proceeds in the following way.
We start by
constructing the symmetry group $\grp$ from the given generator matrices.
The relevant vectors $\normals$ and vertices $\verts$ are then used to
construct the \class{Facet} objects (step 3 of Sec.~\ref{sub:naive}).
Lines
\ref{ConstructFaceHierarchy:newFace}--\ref{ConstructFaceHierarchy:newFaceDone}
create the $n$-face $F_n$.
On line~\ref{ConstructFaceHierarchy:initResultList}, we
initialize the result list $\scrF$, which will contain sets $\scrF_d$ of
$d$-faces for each dimension $d=0, \ldots, n$.
At this point, only the facets ($d=n-1$) and the cell itself ($d=n$) are known.
The loop on line~\ref{ConstructFaceHierarchy:loop}
then iterates from the $(n-2)$-faces down to the $0$-faces.
In each dimension $d$, we classify the parents into equivalence classes,
which arbitrarily selects one representative face in each class.
Then in line~\ref{ConstructFaceHierarchy:intersect},
we intersect each of these representative faces with
all children of one of its (representative) parents
using Alg.~\ref{alg:IntersectFaces}.
As we prove below, this will create all child faces of those representatives,
which are stored in $\scrF_d$.

\begin{alg}
    \caption{\raggedright
        Create child faces of representatives.
    }\label{alg:IntersectFaces}
    \begin{algorithmic}[1]
        \Procedure{IntersectFaces}{{\em faces}}
            \State {\em children} $\gets$ new empty Dictionary
            \ForAll{$P_1 \in$ \Call{Reps}{{\em faces}}}
                    \label{IntersectFaces:reps-loop}
                    \CommentInline{Outer loop over representatives}
                \ForAll{$P_2 \in$ \Call{RelevantSiblings}{$P_1$}}
                        \CommentInline{Inner loop over faces to intersect with}
                    \State $F$ $\gets$ NULL
                        \CommentInline{Storage for a potential child face}
                    \State $\verts \gets P_1$.vertices $\cap\ P_2$.vertices
                        \CommentInline{Intersect the sets of vertices}
                    \If{$\verts \in$ {\em children}}
                            \CommentInline{Child is already known}
                        \State $F \gets$ {\em children}$[\verts]$
                            \CommentInline{Get the face object}
                        \State add $P_1, P_2$ to $F$.parents
                            \CommentInline{Add new parents}
                    \ElsIf{dimension of $\verts = P_1$.dim${}-1$}
                            \CommentInline{Affine space has the correct dimension}
                        \State $F \gets$ new Face
                            \CommentInline{Create a new \class{Face} object}
                        \State $F$.dim $\gets P_1$.dim${}-1$
                        \State $F$.vertices $\gets \verts$
                        \State $F$.parents $\gets \{P_1, P_2\}$
                        \State {\em children}$[\verts] \gets F$
                            \CommentInline{Store that we know this set of vertices}
                    \EndIf
                    \If{$F \neq$ NULL}
                        \State add $F$ to $P_1$.children and $P_2$.children
                            \CommentInline{Register $F$ as child of the two parents}
                    \EndIf
                \EndFor
            \EndFor
            \State \Return Set of values in {\em children} dictionary
        \EndProcedure
        \EmptyLine
        \Procedure{Reps}{{\em faces}}
                \CommentInline{Collects representatives of already classified faces}
            \State $\scrFrep$ $\gets$ new empty List
            \ForAll{$F \in$ {\em faces}}
                \If{$F$.representative $=$ NULL}
                    \State add $F$ to $\scrFrep$
                \EndIf
            \EndFor
            \State \Return $\scrFrep$
        \EndProcedure
        \EmptyLine
        \Procedure{RelevantSiblings}{$F_d$}
                \CommentInline{Collects children of (any) one representative parent}
            \ForAll{$F_{d+1} \in$ $F_d$.parents}
                \If{$F_{d+1}$.representative $=$ NULL}
                    \State \Return $F_{d+1}$.children
                \EndIf
            \EndFor
            \State {\bfseries error} ``Face has no representative parent''
                \CommentInline{All $F_{d<n}$ have a rep.\ parent $\Rightarrow$ this is never reached}
        \EndProcedure
    \end{algorithmic}
\end{alg}

\begin{alg}
    \caption{\raggedright
        Classify a set of faces (mutates {\em faces}).
    }\label{alg:ClassifyFaces}
    \begin{algorithmic}[1]
        \Procedure{ClassifyFaces}{{\em faces}, $\grp$, {\em facets}}
            \State {\em face\_lists} $\gets$ \Call{SplitByFingerprint}{{\em faces}, {\em facets}}
                \label{ClassifyFaces:split}
            \ForAll{$\scrF \in$ {\em face\_lists}}
                    \label{ClassifyFaces:outer-loop}
                    \CommentInline{$\scrF$ is a list of faces with the same fingerprint}
                \State $\scrFrep$ $\gets$ new empty List
                    \CommentInline{This will contain the representatives of all classes found in $\scrF$ thus far}
                \ForAll{$F \in \scrF$}
                    \ForAll{$\Frep \in$ $\scrFrep$}
                        \State $g$ $\gets$ \Call{FindTransformation}{$\Frep$, $F$}
                            \label{ClassifyFaces:equiv-check}
                        \If{$g \neq$ NULL}
                                \CommentInline{$F \sim \Frep$ and $F = g\Frep$}
                            \State $F$.representative $\gets$ $\Frep$
                            \State $F$.transformation $\gets$ g
                            \State {\bfseries break}
                                \CommentInline{No need to check other $\Frep$}
                        \EndIf
                    \EndFor
                    \If{$F$.representative $=$ NULL}
                        \State add $F$ to $\scrFrep$
                            \CommentInline{$F$ is in a new class; mark it as representative}
                        \State $F$.class\_id $\gets$ next available integer
                    \EndIf
                \EndFor
            \EndFor
        \EndProcedure
        \EmptyLine
        \Procedure{SplitByFingerprint}{{\em faces}, {\em facets}}
            \State {\em faces\_by\_fingerprint} $\gets$ new empty Dictionary
            \ForAll{$F \in$ {\em faces}}
                \State $\normals$ $\gets$ \Call{GetNormals}{$F$, {\em facets}}
                \State {\em fp} $\gets$ \Call{PerClassCount}{$F$.vertices $\cup$ $\normals$}
                    \label{SplitByFingerprint:fp}
                    \CommentInline{Calculate a fingerprint {\em fp} (see Sec.~\ref{sub:face-equivalence})}
                \If{{\em fp} $\notin$ {\em faces\_by\_fingerprint}}
                    \CommentInline{This fingerprint is new}
                    \State {\em faces\_by\_fingerprint}$[\text{{\em fp}}]$ $\gets$ [$F$]
                        \CommentInline{Initialize a list for this new fingerprint}
                \Else
                    \State append $F$ to {\em faces\_by\_fingerprint}$[\text{{\em fp}}]$
                \EndIf
            \EndFor
            \State \Return List of values in {\em faces\_by\_fingerprint} dictionary
        \EndProcedure
    \end{algorithmic}
\end{alg}

The classification itself
includes identifying equivalences and finding transformations between
equivalent faces. It
is carried out by the procedure
\proc{ClassifyFaces} shown in Alg.~\ref{alg:ClassifyFaces}.
As a first step,
in line~\ref{ClassifyFaces:split}
it splits the given list of {\em faces} into multiple smaller
lists, each containing only faces that have the same {\em fingerprint}.
Fingerprints will be discussed in more detail in
Sec.~\ref{sub:face-equivalence}.
For now, it is sufficient to know that faces with different fingerprints are
guaranteed to be inequivalent under $\grp$.

A very valuable property of the loop in
line~\ref{ClassifyFaces:outer-loop}
is that each iteration is fully independent of any other iteration.
In particular, no change is made to the data structure outside the faces in the
sub-list $\scrF$ of the current iteration.
Problems with this property are called ``embarrassingly parallel'' since they
can in principle be run in parallel across many cores or even computers in a
cluster.

The procedure \proc{FindTransformation} used in
line~\ref{ClassifyFaces:equiv-check}
is responsible for evaluating whether two faces $F$ and $F'$ are equivalent
under $\grp$.
If they are, it returns one of the group elements (there may be multiple)
such that $F' = g F$.
This highly nontrivial task is discussed in detail in
Sec.~\ref{sub:find-transform},
where an algorithmic solution is given.

\subsubsection{Proving important properties}
\label{sub:faces:proofs}

As mentioned above, it is important that Alg.~\ref{alg:ConstructFaces}
produces results having the properties
\ref{enum:all-children} and \ref{enum:all-classes}
(see Sec.~\ref{sub:faces:overview}), which we will now prove.
We start with \ref{enum:all-children}, i.e., that
Alg.~\ref{alg:ConstructFaces} yields {\em all} children of each representative
face.

We argue by recursion.
First, \ref{enum:all-children} holds trivially in dimension
$d=n$, since all children of the (only)
$n$-face $F_n$ are generated.
Recall that these are precisely the facets, which are all constructed
in line~\ref{ConstructFaceHierarchy:facets} of Alg.~\ref{alg:ConstructFaces}.
Next, assume it is true in dimension $d+1$:
the set of constructed faces of dimension $d$ includes all
children of all representative $(d+1)$-faces.
Within this set of $d$-faces, Alg.~\ref{alg:ConstructFaces} arbitrarily selects one
representative per subset of equivalent $d$-faces.
Let $F$ be one such representative $d$-face.
By construction, $F$ has at least one parent $P$ that is a
representative face.
By assumption, Alg.~\ref{alg:ConstructFaces} includes all
children of $P$ among the faces of dimension $d$.
However, it is clear that all child faces of a face $F$ may be constructed by
picking {\em any} parent $P$ of $F$ and intersecting $F$ with all
{\em other} children of $P$.
It follows immediately that
all children of $F$ are constructed.

We continue with the proof of \ref{enum:all-classes}, which
states that the set $\scrFrep_d \defeq$ \proc{Reps}$(\scrF_d)$
of representatives
of the set of $d$-faces $\scrF_d$ constructed by Alg.~\ref{alg:ConstructFaces}
is complete:
the union of the orbits
\begin{equation}
    \bigcup_{\Frep_d \in \scrFrep_d} \grp \Frep_d
\end{equation}
is the full set of $d$-faces of the Voronoi cell $\Omega$.

Here, the proof is recursive and by contradiction.  Assume that the
union provides a complete face set in dimension $d+1$, but that the
Voronoi cell includes a $d$-face $F'$ which is {\em not}
constructed by Alg.~\ref{alg:ConstructFaces} and is also {\em not} equivalent to one
of the representative $d$-faces which is constructed.  Since $F'$ is a
$d$-face, it must have a parent, which by assumption can be written as
$P = g\Frep_{d+1}$, where $g$ is a group element and $\Frep_{d+1}$
a representative $(d+1)$-face.  In this case, by symmetry, $g^{-1} P$
must have $g^{-1} F'$ as a child.  Since $g^{-1} P = \Frep_{d+1}$ is
a representative face, and Alg.~\ref{alg:ConstructFaces} constructs all children of
representative faces, then $g^{-1} F'$ must have been constructed.
But this contradicts our assumption, since $g^{-1} F'$ is in the same
orbit as $F'$ and hence equivalent to it.

\subsubsection{Iterated classification}
\label{sub:faces:iterated}

The classification step in Alg.~\ref{alg:ConstructFaces} can be further
optimized as follows.
For a given set of faces, one may perform the classification via
\proc{ClassifyFaces}({\em faces}, $\setU$, {\em facets})
using subgroups $\setU \subset \grp$.
More precisely, let
\begin{equation}\label{eq:subgroups-U}
    \setU_1, \setU_2, \ldots, \setU_N
\end{equation}
be a list of proper subgroups of $\grp$.
Generally, we shall order them such that $|\setU_i| \leq |\setU_{i+1}|$
but without implying inclusion relations.

The classification using the subgroups at the beginning of this list is
usually much faster than with the subgroups at the end or with the full group $\grp$.
We therefore produce classes of faces by starting the classification with
$\setU_1$ and working up to $\setU_N$ followed by $\grp$.
In each step, only one representative of the previous classification is
considered.
If two of the representatives turn out to be equivalent, then the whole sets
they represent can be merged.

Note that it is possible for faces in a single class under $\setU_i$ to
be inequivalent under $\setU_{i+1}$.
This is not a problem, as equivalence is clearly restored under $\grp$.

The choice of subgroups may have a large influence on the resulting
computational cost.
In practice, it turned out to be sufficient to use a subset of stabilizers of
the facets and $(n-2)$-dimensional faces.\footnote{%
    We constructed stabilizers for the representatives of the $(n-1)$- and
    $(n-2)$-faces.
    For laminated \Ktwelve, we used only some of these stabilizers.
    This is explained in more detail in Sec.~\ref{sub:laminatedK12}.
}
The stabilizer of an element $X$ (for example a face or a vector) acted on by $\grp$ is the set
\begin{equation}\label{eq:stabilizer}
    \Stab(X) \defeq \left\{ g \in \grp : g X = X \right\} \;.
\end{equation}

It is easy to construct such a stabilizer using \lib{GAP}.
For individual vectors, on which the matrix group acts directly, one may use
the pre-defined action
``OnPoints''
\begin{verbatim}
    gap> Stabilizer(G, x, OnPoints);
\end{verbatim}
where \verb+G+ is the matrix group and \verb+x+ the vector.
This is shown here with \lib{GAP} syntax and is translated accordingly when
used in Python with \lib{gappy}.
To compute the stabilizer of a face, we take a set of vectors defining the
face (see Sec.~\ref{sub:face-equivalence}) and use the pre-defined action
``OnSets''.
For example, for a face defined by three vectors $\vc x_1, \vc x_2, \vc x_3$,
one may use
\begin{verbatim}
    gap> vectors = AsSet([x1, x2, x3]);
    gap> Stabilizer(G, vectors, OnSets);
\end{verbatim}

The next section discusses our approach to evaluating the equivalence of
two faces.

\subsection{Evaluating the equivalence of faces}
\label{sub:face-equivalence}

The above strategy of constructing the hierarchy of faces of
$\Omega$ hinges on an efficient way to evaluate if two faces are equivalent.
In Alg.~\ref{alg:ClassifyFaces}, line~\ref{ClassifyFaces:equiv-check},
this is performed by the call to \proc{FindTransformation}.
The present section presents our approach to solving this problem.
We start in Sec.~\ref{sub:face-eq-vertices} and
\ref{sub:face-eq-relevant-vectors} by discussing two possible sets of vectors
one may use, followed by our decision criterion in Sec.~\ref{sub:defining-vectors}.
Recall that in Alg.~\ref{alg:ClassifyFaces}, the equivalence test is only
performed with faces having the same fingerprint.
Since the specific fingerprint we use is important for our algorithm, it is
presented next in Sec.~\ref{sub:fingerprints}
before we introduce our method (Sec.~\ref{sub:find-transform}) and
the algorithm itself (Sec.~\ref{sub:find-transform-alg}).
Finally, Sec.~\ref{sub:find-transform-formulas} solves the problem of
determining {\em all} transformations in $\grp$ taking one vector into another,
which is a key ingredient in our method.

\subsubsection{Equivalence using vertices}
\label{sub:face-eq-vertices}

We first identify the quantities one may use to evaluate if two faces $F_d$
and $F_d'$ are equivalent.
Recall that $F_d \sim F_d'$ if and only if there exists a $g \in \grp$ such
that $F_d' = g F_d$.
From \eqnref{eq:g-on-polytope}, it is clear that
\begin{equation}\label{eq:face-eq-same-as-verts-eq}
    F_d' = g F_d
    \quad\Longleftrightarrow\quad
    \verts(F_d') = g \verts(F_d) \;,
\end{equation}
that is, we can use the sets of vertices of the two faces.

For faces in higher dimensions, however, the number of vertices can become too
large for computationally evaluating equivalence.
For example, the $10$-faces of \Ktwelve contain between $104$ and $6\,978$
vertices each.

\subsubsection{Equivalence using relevant vectors}
\label{sub:face-eq-relevant-vectors}

An alternative to the vertices
is the set of relevant vectors of the
facets within which a face lies.
For a $d$-face $F_d$, where $d < n$, this set is defined by
\begin{equation}\label{eq:normals-of-face}
    \normals(F_d) \defeq \left\{
        \vc n_i \in \normals(\Omega)
            : F_d \subset E_i
    \right\} \;,
\end{equation}
where $\normals(\Omega)$ is the set of all relevant vectors\footnote{
    Note that for $d=n$, the only face is the Voronoi cell itself,
    $F_n = \Omega$, and so \eqnref{eq:normals-of-face} would result in the
    empty set. We hence restrict this definition to $d < n$ and {\em define}
    $\normals(F_n)$ as the full set of all relevant vectors.
}
and $E_i$ are the planes \eqref{eq:facet-planes} the corresponding facets lie in.
The vectors $\vc n \in \normals(F_d)$ are all orthogonal to $F_d$.
In fact, since all faces of dimension $d<n$ are ultimately obtained by repeated
intersections of facets, \eqnref{eq:normals-of-face} collects the
relevant vectors of the facets whose intersection yields $F_d$, i.e.,
\begin{equation}\label{eq:Fd-is-intersection}
    F_d = \bigcap_{\vc n \in \normals(F_d)} F^{\vc n}
    \;.
\end{equation}
Here, $F^{\vc n}$ is the facet belonging the relevant vector
$\vc n \in \normals(\Omega)$.

The number of vectors in $\normals(F_d)$ depends on the dimension $d$ of the
face.
To see this, consider $k$ linearly independent vectors
$\vc n_i \in \normals(\Omega)$.
The intersection of the corresponding $E_i$ is an affine space of dimension
$n-k$.
This means that $\normals(F_d)$ will contain at least $k=n-d$ vectors,
since $n-k=d$.
$\normals(F_d)$ may contain more elements,
as can be seen in Fig.~\ref{fig:intersection-too-low-dim}.
For the face $F_0^1$, we have $n=3$ and $d=0$,
but the set $\normals(F_0^1)$ contains the four normals of the facets
$F_2^1, \ldots, F_2^4$.

An important observation for us is that
\begin{equation}\label{eq:face-eq-same-as-normals-eq}
    F_d' = g F_d
    \quad\Longleftrightarrow\quad
    \normals(F_d') = g \normals(F_d) \;.
\end{equation}
To prove this, first note that by \eqref{eq:facet-planes}, we
have
\begin{align}
    \notag
    g E_i ={}& \left\{ g \vc x \in \R^n : 2\,\vc x \cdot \vc n_i = \|\vc n_i\|^2 \right\} \\
        \notag
        ={}& \left\{ \vc x \in \R^n : 2 (g^{-1} \vc x) \cdot \vc n_i = \|\vc n_i\|^2 \right\} \\
        \notag
        ={}& \left\{ \vc x \in \R^n : 2 \vc x \cdot (g \vc n_i) = \|\vc n_i\|^2 \right\} \\
        \label{eq:gE}
        ={}& E_{i'}
        \;,
\end{align}
where $E_{i'}$ is the plane corresponding to the transformed normal
$g \vc n_i$.
Let now $F_d$ be a $d$-face, where $d < n$, and $g \in \grp$.
Then, since
$F_d \subset E_i \Longleftrightarrow g F_d \subset g E_i$
and using the definition \eqref{eq:normals-of-face} and \eqref{eq:gE},
we get $\normals(g F_d) = g \normals(F_d)$.
For the other direction, let $F_d$ and $F_d'$ be two $d$-faces.
Since $g F^{\vc n} = F^{g \vc n}$,
we have with \eqref{eq:Fd-is-intersection}
\begin{equation}\label{eq:gFd-via-intersection}
    g F_d = \bigcap_{\vc n \in \normals(F_d)} F^{g \vc n}
        = \bigcap_{\vc n \in g \normals(F_d)} F^{\vc n}
    \;.
\end{equation}
If now
$g \normals(F_d) = \normals(F_d')$,
this implies $g F_d = F_d'$.

Together, \eqref{eq:face-eq-same-as-verts-eq} and
\eqref{eq:face-eq-same-as-normals-eq}
mean that we are free to use either $\verts(F_d)$ or $\normals(F_d)$ to
evaluate equivalence.
In the example of \Ktwelve, the set \eqref{eq:normals-of-face} has exactly $2$
elements for all $10$-faces.

We use Alg.~\ref{alg:GetNormals} to obtain $\normals(F_d)$ in our code.
This is made particularly efficient by our data structure.
We simply iterate over all facets and check if the \field{vertices} field of
$F_d$ is a subset of the \field{vertices} field of the respective facet
(line~\ref{GetNormals:if}).
Recall from Fig.~\ref{fig:data-structure}
that facets are instances of the subclass \class{Facet} and thus store
their relevant vector in the \field{normal} field.
Since $\normals(F_d)$ is used multiple times in our algorithm, we store the
normals in the \field{facet\_normals} field for later re-use.

\begin{alg}
    \caption{\raggedright
        Compute the set $\normals(F_d)$.
    }\label{alg:GetNormals}
    \begin{algorithmic}[1]
        \Procedure{GetNormals}{$F_d$, {\em facets}}
            \If{$F_d$.facet\_normals $\neq$ NULL}
                \State \Return $F_d$.facet\_normals
            \EndIf
            \State $\normals$ $\gets$ new empty Set
            \ForAll{$F_{n-1} \in$ {\em facets}}
                \If{$F_d$.vertices $\subseteq F_{n-1}$.vertices}
                        \label{GetNormals:if}
                    \State $\normals \gets$ $\normals \cup \{ F_{n-1}\text{.normal} \}$
                        \CommentInline{$F_d$ is a subface of $F_{n-1}$; collect its normal}
                \EndIf
            \EndFor
            \State $F_d$.facet\_normals $\gets$ $\normals$
            \State \Return $\normals$
        \EndProcedure
    \end{algorithmic}
\end{alg}

We remark that Alg.~\ref{alg:GetNormals} could have been implemented by
recursively collecting all parent faces up to dimension $n-1$.
However, this does not work in our case since the information about a face's
parents is not guaranteed to be complete and so some parents may be missed.

The next steps require that the
fields \field{representative}, \field{transformation} and \field{class\_id}
of the
relevant vectors
$\normals(\Omega)$ have been populated.
As described in Sec.~\ref{sub:vertices}, the vertices have already been
classified as part of the finding process, i.e., these
fields are fully populated.
For the much smaller number of relevant vectors, we will assume
that this has been done in a similar way.

\subsubsection{Deciding which set of vectors to use}
\label{sub:defining-vectors}

We can now construct a {\em defining set} $\defD(F_d)$ for each face, which will
be used for evaluating equivalence.
A simple choice for $\defD(F_d)$ would be the smaller of the two sets $\verts(F_d)$
and $\normals(F_d)$.
However, for reasons that will become clear below, we instead take the
following approach.
First, both sets $\verts(F_d)$ and $\normals(F_d)$ are sorted such that
equivalent vectors are adjacent to each other.
We then count the number of permutations that keep equivalent vectors adjacent
and the classes in the same order.
The set with the smaller number of these
permutations
is chosen as $\defD(F_d)$.
In case both sets have the same number of those permutations, we choose the
smaller set, or, for definiteness,
$\normals(F_d)$ if that number is also equal.
We do not present this logic in a formal algorithm.
It will be used later in Alg.~\ref{alg:FindTransformation},
line~\ref{PartitionDefiningSet:def}.

The
criterion for the choice of $\defD(F_d)$ is based purely on invariants
under $\grp$:
two equivalent faces necessarily have the same number of vertices and
relevant vectors in each class.
In general, $\defD(F_d) = \verts(F_d)$ for small $d$ and
$\defD(F_d) = \normals(F_d)$ for large $d$, but it is not only a function of
$d$.
For two inequivalent $d$-faces $F_d$ and $F_d'$, it may happen that
$\defD(F_d) = \normals(F_d)$ but $\defD(F_d') = \verts(F_d')$.

\subsubsection{Fingerprints of faces}
\label{sub:fingerprints}

Before deciding if $F_d' \sim F_d$, it is possible to perform
several inexpensive tests that can show if equivalence is excluded.
Any property of a face that is invariant under $\grp$ can be used for this
purpose.
We will call such a property a {\em fingerprint} of a face.

With our data structure, several fingerprints of a face $F_d$ are easily
calculated.
Obvious ones are the total number of vertices $|\verts(F_d)|$ and of
facet normals $|\normals(F_d)|$.

The number of children of a face is another fingerprint.
However, it is not available in our algorithm as the classification is
performed prior to constructing the child faces.

An additional fingerprint is the per-class count of elements in $\verts(F_d)$
and $\normals(F_d)$, which is calculated by Alg.~\ref{alg:PerClassCount}.
The per-class count
contains all the information needed to evaluate the criterion
for the choice of $\defD(F_d)$ described at the end of
Sec.~\ref{sub:defining-vectors}.
In fact, we use the per-class count of the union
$\verts(F_d) \cup \normals(F_d)$
in line~\ref{SplitByFingerprint:fp} of Alg.~\ref{alg:ClassifyFaces}
to partition the faces as a first step in their classification.

\begin{alg}
    \caption{\raggedright
        Compute the per-class count of a set of vectors.
    }\label{alg:PerClassCount}
    \begin{algorithmic}[1]
        \Procedure{PerClassCount}{{\em vectors}}
            \State {\em pcc} $\gets$ new empty Dictionary
            \ForAll{$\vc v \in$ {\em vectors}}
                \State {\em id} $\gets$ $\vc v$.class\_id
                    \CommentInline{Get this vector's \field{class\_id}}
                \If{{\em id} $=$ NULL}
                    \State {\em id} $\gets$ $\vc v$.representative.class\_id
                \EndIf
                \If{{\em id} $\in$ {\em pcc}}
                        \CommentInline{Check if we have a counter for this {\em id}}
                    \State {\em pcc}[{\em id}] $\gets$ {\em pcc}[{\em id}] ${}+ 1$
                        \CommentInline{Yes we do; increment it}
                \Else
                        \CommentInline{No we do not; initialize a new one}
                    \State {\em pcc}[{\em id}] $\gets$ 1
                \EndIf
            \EndFor
            \State {\em result} $\gets$ list of $(\text{{\em key}}, \text{{\em value}})$ pairs in {\em pcc} sorted by {\em key}
            \State \Return {\em result}
        \EndProcedure
    \end{algorithmic}
\end{alg}

\subsubsection{A strategy for finding a transformation}
\label{sub:find-transform}

Our goal now is to determine if $\defD(F_d')$ and $\defD(F_d)$ are equivalent,
and, in case they are, to find a symmetry $g \in \grp$ such that
$\defD(F_d') = g \defD(F_d)$.
We will assume that the two faces $F_d$ and $F_d'$ have passed the fingerprint
test discussed above.
This means that the defining sets
$\defD(F_d)$ and $\defD(F_d')$
have an equal number of vectors per equivalence class
and we can form $N$ pairs of individually equivalent vectors.
Here, $N \defeq |\defD(F_d)| = |\defD(F_d')|$ is the number of vectors in
each of the defining sets.

\begin{figure}\centering
    \includegraphics{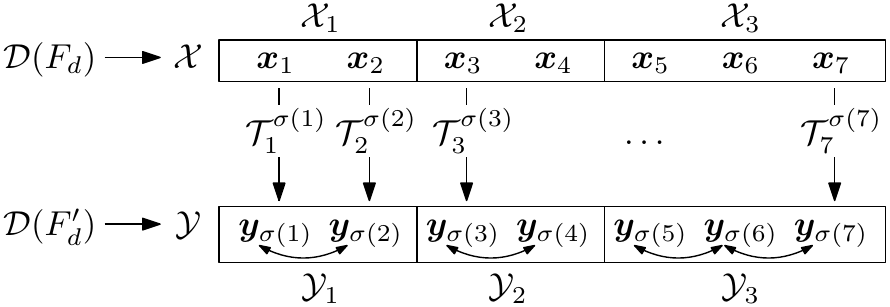}%
    \caption[]{\label{fig:partitions}%
        Scheme of the construction of transformations between
        the vectors $\vc x_i$ and $\vc y_{\sigma(i)}$
        of the defining vectors of $F_d$ and $F_d'$.
        The permutation $\sigma$ switches only the positions of vectors
        within their respective subset $\YY_j$ of equivalent vectors.
    }
\end{figure}

Our method is illustrated in
Fig.~\ref{fig:partitions}.
The sets $\defD(F_d)$ and $\defD(F_d')$ are each split into
subsets of equivalent vectors.
In our algorithms, the partition $\XX$ of $\defD(F_d)$
is represented as a list of lists $\XX_j$.
For a fixed $j$, the vectors $\vc x \in \XX_j$ are all equivalent.
We have in this example $N=7$, $|\XX| = 3$ and $|\XX_j| = 2, 2, 3$ for $j = 1, 2, 3$,
respectively.
Similarly, $\YY$ is a partition of $\defD(F_d')$ with subsets $\YY_j$.
We first order the subsets $\XX_j$ and $\YY_j$ by the \field{class\_id} field of the contained
vectors.
Since the fingerprint test has passed,
the vectors in $\XX_j$ and those in $\YY_j$ are in the same
class for each $j$.
We further order the subsets such that
\begin{equation}\label{eq:partition-ordering}
    |\XX_j| = |\YY_j|
    \leq
    |\XX_{j+1}| = |\YY_{j+1}|
    \;.
\end{equation}
This is done with a stable sorting algorithm so that for each $j$ the vectors in
$\XX_j$ and $\YY_j$ remain in the same class, even when
$|\XX_j| = |\XX_{j+1}|$.

The main idea of our algorithm is now as follows.
Let $\sigma$ be a permutation of $1, \ldots, N$, which keeps the pairs
equivalent, $\vc x_i \sim \vc y_{\sigma(i)}$ for all $i = 1, \ldots, N$.
In other words, $\sigma$ permutes the vectors $\vc y_i$ only within the
individual subsets $\YY_j$,
which is indicated by the arrows inside the $\YY$ row
in Fig.~\ref{fig:partitions}.
The single permutation $\sigma$ is equivalent to a set
$\{\sigma_1,\ldots,\sigma_{|\YY|}\}$ of permutations acting on the individual subsets $\YY_j$.
Let further
\begin{equation}\label{eq:all-transforms}
    \setT_i^k \defeq \{ g \in \grp : \vc y_k = g \vc x_i \}
\end{equation}
be the set of all group elements taking the single vector $\vc x_i$ into
the single vector $\vc y_k$.
Clearly, if there exists a permutation $\sigma$ such that
\begin{equation}\label{eq:intersection-of-Ti}
    \bigcap_{i=1}^N \setT_i^{\sigma(i)}
        \neq \anemptyset
    \;,
\end{equation}
then $F_d \sim F_d'$ and any group element $g$ in this intersection takes
$F_d$ into $F_d'$.
If the intersection is empty for all permutations,
then $F_d$ and $F_d'$ are inequivalent.

One property of this method that may become a limitation
is the scaling with the factorial of the sizes
of the subsets $\YY_j$.
By permuting only within the subsets, the number of permutations to check
is reduced from $N!$ to
\begin{equation}\label{eq:total-permutations}
    \prod_{j} |\YY_j|! \;.
\end{equation}
We argue below that this number is a worst case only and we discuss several
means by which we can often avoid most of these permutations.

Despite this, our algorithm may fail if the subset sizes $|\YY_j|$ become
too large and thus too many permutations need to be checked.
For such cases, one may switch to a different algorithm.
Alternatives include the \lib{GAP} function
``RepresentativeAction'', which would replace the call to
\proc{FindTransformation} in Alg.~\ref{alg:ConstructFaces}.
However, in the cases we encountered for the $12$ and $13$-dimensional
lattices discussed in Sec.~\ref{sec:results},
the method described here was superior.

\subsubsection{An algorithm for finding a transformation}
\label{sub:find-transform-alg}

\begin{alg}[!t]
    \caption{\raggedright
        Evaluate if $F_d \sim F_d'$.
        Return a group element $g \in \grp$ such that $F_d' = g F_d$ if
        $F_d \sim F_d'$, else NULL.
    }\label{alg:FindTransformation}
    \begin{algorithmic}[1]
        \Procedure{FindTransformation}{$F_d$, $F_d'$}
            \State $\XX \gets$ \Call{PartitionDefiningSet}{$F_d$}
                \label{FindTransformation:partitioning1}
                \CommentInline{Split vectors by class and order them as in Fig.~\ref{fig:partitions}}
            \State $\YY \gets$ \Call{PartitionDefiningSet}{$F_d'$}
                \label{FindTransformation:partitioning2}
            \State $\setT \gets$ \Call{FilterTransforms}{$\XX$, $\YY$, $\grp$, 1}
                \label{FindTransformation:filter}
                \CommentInline{Start recursion with full group $\grp$ and
                    subset pair $(\XX_1, \YY_1)$}
            \If{$\setT \neq \anemptyset$}
                    \CommentInline{The faces are equivalent}
                \State \Return one arbitrary element $g \in \setT$
            \Else
                    \CommentInline{The faces are {\em not} equivalent}
                \State \Return NULL
            \EndIf
        \EndProcedure
        \EmptyLine
        \Procedure{FilterTransforms}{$\XX$, $\YY$, $\setP_{j-1}$, $j$}
            \If{$j > |\XX|$} \CommentInline{Recursion stopping criterion}
                \State \Return $\setP_{j-1}$
            \EndIf
            \ForAll{$\sigma_j \in \Perm(1,\ldots,|\XX_j|)$}
                    \label{FilterTransforms:loop}
                    \CommentInline{Iterate over all permutations of the $j$th subset}
                \CommentLineEmpty[0]{Intersect $\setP_{j-1}$ with all transformations of all
                    vector pairs in this pair of subsets:}
                \State $\setP_j \gets $\Call{TransformsOfSubsets}{$\XX_j$, $\sigma_j \YY_j$, $\setP_{j-1}$}
                \If{$\setP_j \neq \anemptyset$} \label{FilterTransforms:if-U-empty}
                    \State $\setP_j \gets $\Call{FilterTransforms}{$\XX$, $\YY$, $\setP_j$, $j + 1$}
                        \label{FilterTransforms:recurse}
                        \CommentInline{Recurse to next pair of subsets $(\XX_{j+1}, \YY_{j+1})$}
                    \If{$\setP_j \neq \anemptyset$}
                        \CommentLineEmpty[0]{{\em All} following subsets have a permutation admitting the transformations left in $\setP_j$.}
                        \CommentLineEmpty[0]{This means $F \sim F'$ and we are done:}
                        \State \Return $\setP_j$
                    \EndIf
                    \CommentLineEmpty[0]{Subsequent subsets are not compatible with any $g \in \setP_j$ under any permutation.}
                    \CommentLineEmpty[0]{Try next permutation of this pair of subsets.}
                \Else
                    \CommentLineEmpty[0]{No transformation left. Try next permutation of $\YY_j$.}
                \EndIf
            \EndFor
            \State \Return $\anemptyset$
                \CommentInline{All permutations of $\YY_j$ exhausted; try next permutation of $\YY_{j-1}$}
        \EndProcedure
        \EmptyLine
        \Procedure{TransformsOfSubsets}{$\XX_j$, $\YY_j$, $\setP_{j-1}$}
            \State $\setP_j \gets \setP_{j-1}$
            \ForAll{$1 \leq k \leq |\XX_j|$}
                    \CommentInline{Iterate over all pairs of vectors in $\XX_j$ and $\YY_j$}
                \State $\vc x$ $\gets$ $k$th vector of the list $\XX_j$
                \State $\vc y$ $\gets$ $k$th vector of the list $\YY_j$
                \State $\setT_{xy} \gets$ \eqnref{eq:cosets}
                    \label{TransformationsOfPartition:Txy}
                    \CommentInline{Construct all transformations of the $k$th pair}
                \State $\setP_j \gets \setP_j \cap \setT_{xy}$
                    \CommentInline{Intersect to filter the pool $\setP_j$}
                \If{$\setP_j = \anemptyset$} \CommentInline{Nothing left; skip remaining pairs}
                    \State \Return $\anemptyset$
                \EndIf
                \CommentLineEmpty[0]{Pool is not empty, go to next pair.}
            \EndFor
            \State \Return $\setP_j$
        \EndProcedure
        \EmptyLine
        \Procedure{PartitionDefiningSet}{$F$}
            \State $\defD$ $\gets$ defining set of $F$ as in Sec.~\ref{sub:defining-vectors}
                \label{PartitionDefiningSet:def}
            \State $\XX$ $\gets$ partition $\defD$ based on the \field{class\_id} of the vectors
                \CommentInline{We store $\XX$ as list $[\XX_1, \XX_2,\ldots]$ of lists $\XX_j$}
            \State sort $\XX$ based on the \field{class\_id} of the vectors in the $\XX_j$
            \State stably sort $\XX$ based on the sizes $|\XX_j|$
            \State \Return $\XX$
        \EndProcedure
    \end{algorithmic}
\end{alg}

In this section, we describe our implementation of the method outlined in
Sec.~\ref{sub:find-transform}.

A key point is to keep track of a pool of
{\em remaining transformations}, defined as follows.
Let $\setP_0 = \grp$ be the full group. We set
\begin{equation}\label{eq:pool}
    \setP_{j}^{\sigma_1, \ldots, \sigma_j}
        \defeq \bigcap_{i=1}^L \setT_i^{\sigma(i)} \;,
\end{equation}
where $L = \sum_{k=1}^j |\XX_k|$ is the number of vectors up to and including
the $j$th pair of subsets $(\XX_j, \YY_j)$ and where $\sigma$ corresponds to
the permutation of the first $L$ vectors in $\defD(F_d')$ defined by
the permutations $\sigma_1, \ldots, \sigma_j$ of the first $j$ subsets $\YY_k$.
For brevity, we will drop the explicit dependency of the pool on the
permutations $\sigma_1, \ldots, \sigma_j$ and write $\setP_j$.

Our method now recursively goes through all pairs $(\XX_j, \YY_j)$,
$j=1,\ldots,|\XX|$.
Generally, we move down to the next pair as soon as we find a permutation
$\sigma_j$ such that $\setP_j$ is nonempty.
We move back up one level to try the next permutation $\sigma_{j-1}$ if none are found
in the current level.

To go from $j$ to $j+1$ we first initialize
$\setP_{j+1} \gets \setP_j$.
For each pair $(\vc x_i, \vc y_{\sigma(i)})$ of vectors
$\vc x_i \in \XX_{j+1}$ and $\vc y_{\sigma(i)} \in \YY_{j+1}$,
we then construct
the set of transformations $\setT_i^{\sigma(i)}$ and update
$\setP_{j+1} \gets \setP_{j+1} \cap \setT_i^{\sigma(i)}$.

Two cases can occur during this process.
First, $\setP_j$ may become empty at some point when going through the pairs of
vectors in the $j$th pair of subsets.
In that case, we iterate over the permutations $\sigma_j$ of $\YY_j$ and each time repeat
this process from the beginning of the $j$th pair of subsets.
This is done until a permutation $\sigma_j$ is found where $\setP_j$ remains nonempty.
If one is found, we go to the next pair $(\XX_{j+1}, \YY_{j+1})$.
If none is found, we go back to the pair $j-1$ and try the next permutation
of $\YY_{j-1}$.
Finally, if the permutations of $\YY_1$ are exhausted, the faces are inequivalent.
The second case is that we end up with a nonempty set of remaining
transformations $\setP_j$ after going through all
pairs of subsets.
This immediately ends the search, since any of its elements takes $F_d$ into $F_d'$.

Alg.~\ref{alg:FindTransformation} implements this idea as follows
(the implicit dependency of the procedures on the full group $\grp$ is
suppressed).
Lines
\ref{FindTransformation:partitioning1} and
\ref{FindTransformation:partitioning2}
construct the partitions $\XX$ and $\YY$ (i.e., the boxes in Fig.~\ref{fig:partitions})
ordered by their size
and such that the vectors in $(X_j, Y_j)$ are equivalent
for each $j$.
We then enter the recursive procedure
\proc{FilterTransforms}.
It takes as arguments the partitions
$\XX$ and $\YY$ as well as the
pool $P_{j-1}$ of remaining transformations and subset index $j$ to work on
next.
The recursion is started in
line~\ref{FindTransformation:filter},
with the full group $\setP_0=\grp$ as pool and $j=1$ to begin with the
first pair of subsets.
Line~\ref{FilterTransforms:recurse} recurses by advancing to the next
pair of subsets.

Note that, in general, not all permutations $\sigma$ need to be checked, not
even those that respect the partition.
First, if the faces are equivalent, then this equivalence can be manifest in
many of the permutations.
On the other hand, even if the faces are inequivalent, in many cases the
algorithm does not need to check all permutations.
To see this, consider again Fig.~\ref{fig:partitions}.
There are two vector pairs in the first subset and
hence only two permutations of $\YY_1$.
If
$\setT_1^1 \cap \setT_2^2$ and $\setT_1^2 \cap \setT_2^1$ are both empty,
then none of the permutations of $\YY_2$ or $\YY_3$ are considered.
This is the reason for ordering the subsets $\XX_j$ and $\YY_j$ by increasing size:
the larger subsets are only looked at if transformations exist for the
smaller ones.

A further significant reduction of the number of permutations is achieved by
the iterated classification described in
Sec.~\ref{sub:faces:iterated}.
Recall that in this method, we use proper subgroups
$\setU_i \subset \grp$
(usually stabilizers of the $(n-1)$- and $(n-2)$-faces)
to evaluate if two faces are equivalent.
This creates subsets of equivalent faces, each of which is represented by only
one face.
This classification is repeated with different subgroups, where in each
iteration we consider only those representatives.
In the last iteration, the full group $\grp$ is used on a now much smaller set
of faces representing the already collected subsets of equivalent faces.

The number of permutations is generally much smaller for the smaller subgroups
$\setU_i$.
The reason is that vectors that are equivalent under $\grp$ may become
inequivalent under $\setU_i$.
Consider the case shown in Fig.~\ref{fig:partitions}.
A subgroup $\setU_i$ may partition $\defD(F_d)$ and $\defD(F_d')$ into, e.g., four or
five subsets $\XX_j$ and $\YY_j$, respectively, which can drastically decrease
the number of permutations \eqref{eq:total-permutations}.

\begin{centeredtable}{%
    Example cases appearing in
    Alg.~\ref{alg:FindTransformation}
    in the construction of laminated \Ktwelve
    (see Sec.~\ref{sub:laminatedK12}).
    Of the about $8.5$ million invocations of {\upshape\proc{\footnotesize FindTransformation}}
    (including invocations with subgroups $\setU_i \subset \grp$),
    the ten with the largest number of potential checks
    \eqnref{eq:total-permutations} are listed.
    From left to right,
    the columns show
    the dimension $d$ of the face,
    the number $|\defD(F_d)|$ of defining vectors,
    the subset sizes $|\XX_j|$,
    the maximum number \eqnref{eq:total-permutations} of permutations to check,
    the number of permutations actually iterated over
    and whether the respective faces are equivalent.
    In the listed cases, the classification was performed with the
    full group $\grp$.
}
    \label{tab:perms-worst-cases}
    \begin{tabular}{*{6}{r}} \hline
        $d$ & $|\defD(F_d)|$ & $(|\XX_1|, \ldots)$ & $\prod_{j} |\XX_j|!$ &  checked   & equiv.\\ \hline
        $5$ &           $14$ &           $(1,6,7)$ &        $3\,628\,800$ & $169\,908$ &    yes\\
        $5$ &           $14$ &           $(1,6,7)$ &        $3\,628\,800$ & $152\,903$ &    yes\\
        $4$ &           $10$ &              $(10)$ &        $3\,628\,800$ & $127\,811$ &    yes\\
        $5$ &           $14$ &           $(1,6,7)$ &        $3\,628\,800$ &  $30\,738$ &    yes\\
        $4$ &           $10$ &              $(10)$ &        $3\,628\,800$ &  $12\,553$ &    yes\\
        $5$ &           $12$ &             $(4,8)$ &           $967\,680$ &  $14\,563$ &    yes\\
        $5$ &           $12$ &             $(4,8)$ &           $967\,680$ &   $9\,953$ &    yes\\
        $5$ &           $12$ &             $(4,8)$ &           $967\,680$ &   $9\,073$ &    yes\\
        $5$ &           $13$ &           $(1,5,7)$ &           $604\,800$ &   $1\,772$ &    yes\\
        $5$ &           $13$ &           $(1,6,6)$ &           $518\,400$ &      $735$ &    yes\\
        \hline
    \end{tabular}
\end{centeredtable}

The maximal values of $|\XX_j|$ across all face classes in all
dimensions are
$6$ for \AEnine and
$10$ for both \Ktwelve and laminated \Ktwelve.
For laminated \Ktwelve, we list in Tab.~\ref{tab:perms-worst-cases} the worst
cases of numbers of permutations \eqnref{eq:total-permutations} occurring in
invocations of \proc{FindTransformation}.
We note that for inequivalent faces, there were four cases where all
$241\,920$ permutations were checked.
On average, in the roughly $8.5$ million invocations,
about $6.1$ of $34.8$ permutations had to be iterated over for equivalent
faces
and about $9.2$ of $23.1$ permutations for inequivalent faces.

\subsubsection{Constructing the full set of transformations between two vectors}
\label{sub:find-transform-formulas}

The only remaining calculation to describe is the construction of the full set
of transformations $\setT_i^{\sigma(i)}$ taking $\vc x_i$ into
$\vc y_{\sigma(i)}$
for a given $i$.
In Alg.~\ref{alg:FindTransformation}, this is done in
line~\ref{TransformationsOfPartition:Txy}, where we changed the notation
for simplicity,
i.e., we consider the pair of vectors $(\vc x, \vc y)$ and the set
$\setT_{xy}$ of all transformations taking $\vc x$ to $\vc y$.
Luckily, since the fields
\field{representative} and \field{transformation}
of each vector object
are populated, we always know one\footnote{
    The field \field{transformation} of a vector $\vc x$ holds the element
    $g_x \in \grp$ taking the \field{representative} $\vc \vrep$ into
    $\vc x$.
    Similarly, we know one $g_y \in \grp$ such that $\vc y = g_y \vc \vrep$.
    Then, one transformation between equivalent vectors
    $\vc x$ and $\vc y$ is
    given by $g_{xy} \defeq g_y g_x^{-1}$, since
    $\vc y = g_y \vc \vrep = g_y g_x^{-1} g_x \vc \vrep = g_{xy} \vc x$.
}
element $g_{xy} \in \grp$ such that $\vc y = g_{xy} \vc x$.
Then $\setT_{xy}$ is given by the left coset
\begin{equation}\label{eq:cosets}
    \setT_{xy} = g_{xy} \Stab(\vc x) \;,
\end{equation}
where $\Stab(\vc x)$ is the stabilizer \eqref{eq:stabilizer} of $\vc x$.


Note that if the stabilizer $\Stab(\vrep)$
of the (arbitrary but fixed) representative
vector $\vrep$ of a given class has
been calculated and stored,
then the stabilizer of any other vector $\vc x = g_x \vrep$ in the same class
can efficiently be generated via conjugation
\begin{equation}\label{eq:other-stabilizers}
    \Stab(\vc x) = g_x \Stab(\vc \vrep)\, g_x^{-1} \;.
\end{equation}


We make use of this fact by caching the stabilizers of the representative
vectors $\vrep$.
The stabilizer of a vector depends not only on the vector itself but
also on the group.
Since we use several groups $\setU_i$ in the iterated classification described
in Sec.~\ref{sub:faces:iterated},
a single vector generally has multiple stabilizers we need to keep
track of.
This is easily achieved by storing them in a dictionary field of each
individual group object (see the \class{Group} class in Fig.~\ref{fig:data-structure}).
The \field{stabilizer\_cache} has the vectors as keys and the stabilizers as
values.
Whenever a stabilizer of a vector $\vc x = g_x \vrep$ is requested from a
group object, we query its cache for the $\vrep$ key.
If it is not found, the stabilizer is calculated (via \lib{GAP}) and stored in
the cache.
Then we calculate $\Stab(\vc x)$ using \eqnref{eq:other-stabilizers}.

\section{Calculating the second moment}
\label{sec:second-moment}

With the face hierarchy built as discussed in the previous section, it is
possible to calculate the second moment tensor \eqref{eq:Uab} of the
Voronoi cell $\Omega$ using explicit recursion relations.
The strategy for our calculations closely follows the method described in
Section~3 of \cite{allen21}, which corrects and extends the
original relations derived in Section~IV-C of \cite{ViterboBiglieri}.

In our case, however, we can greatly reduce the computational cost of
evaluating these relations by making use of the symmetries of the faces.
When a quantity has been computed for a representative in
$\scrFrep_d$, we immediately obtain this quantity for all equivalent faces.
Scalar quantities are equal for all faces within a class, while
vector or tensor quantities can be transformed
using the known group element.
This significantly reduces the number of steps in our calculations.

Let $F_d \in \scrFrep_d$ be the selected representative of one class of
$d$-faces.
Its centroid $\vc c(F_d)$ is defined as the mean of its vertices $\verts(F_d)$,
\begin{equation}\label{eq:centroid}
    \vc c(F_d) \defeq \frac{1}{|\verts(F_d)|} \sum_{\vc v \in \verts(F_d)} \vc v \;,
\end{equation}
and, due to the convexity of $F_d$, the centroid is guaranteed to lie in $F_d$.
The volume of $F_d$ is defined by
\begin{equation}\label{eq:volume-integral}
    \Vol(F_d) \defeq \int_{F_d} d^d \vc x \;.
\end{equation}
A recursion relation for the volume is
\begin{equation}\label{eq:volume-recursive}
    \Vol(F_d) = \frac{1}{d} \sum_{F \in \children(F_d)} h_{\vc c}(F) \Vol(F) \;,
\end{equation}
where $\children(F_d) \subseteq \scrF_{d-1}$ is the set of all child faces of $F_d$
and $h_{\vc c}(F)$ is the height of the parent's centroid, $\vc c(F_d)$, above the plane containing
its child face $F$.

The height $h_{\vc c}(F)$ is calculated in either of
two ways described in \cite{allen21}.
The first is to project the difference
$\Delta \vc c_F \defeq \vc c(F) - \vc c(F_d)$
onto the space orthogonal to the child face $F$
and then calculating the norm.
This requires the orthogonalization and normalization of a
linearly independent set of, e.g.,
relevant vectors $\normals(F)$ orthogonal to $F$.

The second method of computing $h_{\vc c}(F)$ is to construct any set of vectors
$\vc v_1, \vc v_2, \ldots, \vc v_{d-1}$,
spanning the $(d-1)$-dimensional plane the child $F$ lies in.\footnote{
    This set of spanning vectors can be obtained recursively, starting with the
    $1$-faces and taking the vector connecting its two vertices.
    Then, in dimensions $k>1$, we take the spanning vectors of an arbitrary child face
    and append the vector connecting the centroid of the $k$-face to the centroid
    of the child.
}
Then we have
\begin{equation}\label{eq:h-using-gram-dets}
    h_{\vc c}(F)^2 = \frac{\det\Gram(\vc v_1, \ldots, \vc v_{d-1}, \Delta\vc c_F)}
        {\det\Gram(\vc v_1, \ldots, \vc v_{d-1})}
        \;,
\end{equation}
where $\Gram(\,\cdot\,)$ is the Gram matrix of the vectors in its argument.

As in \cite{allen21}, the decision on which method to use is based on the
dimension $d$ of the face.
For low dimensions, calculating the spanning vectors and Gram determinants
turned out to be faster while in higher dimensions, the projections are less
expensive.
We switch the method usually close to $n/2$, although the precise threshold dimension
seemed not to become a major factor up to $n=13$.

Being a scalar quantity that is invariant under $\grp$, we calculate the volume
once for each face in $\scrFrep_{d-1}$ and subsequently obtain $\Vol(F)$ of
each child $F$ of the $d$-face $F_d$ from its respective representative,
i.e., $\Vol(F) = \Vol(\Frep)$ if $F \sim \Frep \in \scrFrep_{d-1}$.
The recursion starts with the $0$-faces $F_0 \in \scrFrep_0$, which have a $0$-volume of $1$.

Similarly, the barycenter of $F_d$,
\begin{equation}\label{eq:bary}
    \vc b(F_d) \defeq \frac{1}{\Vol(F_d)} \int_{F_d} \vc x\ d^d \vc x \;,
\end{equation}
satisfies a recursion relation \cite[Eq.~(3.5)]{allen21}
\begin{align}
    \label{eq:bary-recursive}
    \vc b(F_d) &= \frac{1}{d+1} \Bigg(
        \vc c(F_d) + \frac{1}{\Vol(F_d)}
        \sum_{F \in \children(F_d)} h_{\vc c}(F) \Vol(F) \vc b(F)
    \Bigg) \;.
\end{align}
Again, the barycenter $\vc b(F)$ of a child face $F$ can be obtained from the
representative of its class.
Since it is a vector quantity, however, it needs to be transformed via
\begin{equation}\label{eq:bary-from-rep}
    \vc b(F) = g \vc b(\Frep) \;,
\end{equation}
where $F = g \Frep$ and $\Frep \in \scrFrep_{d-1}$.
The second moment tensor \eqref{eq:Uab} can be calculated using
\begin{align}
    \label{eq:Uab-recursive}
    \mat U(F_d) =
        \frac{1}{d+2}
        \sum_{F \in \children(F_d)} \! h_{\vc b}(F) \Big[
            \mat U(F) + (\Delta \vc b_F)^T\, \Delta \vc b_F \, \Vol(F)
        \Big] \;.
\end{align}
where $h_{\vc b}(F)$ is the height of the barycenter $\vc b(F_d)$ above the plane of
$F$ and we use the abbreviation
$\Delta \vc b_F \defeq \vc b(F_d) - \vc b(F)$.
The heights $h_{\vc b}(F)$ are calculated in the same way as $h_{\vc c}(F)$.

The second moment tensors $\mat U(F)$ of child faces are obtained from those of
their representative faces as
\begin{equation}\label{eq:Uab-from-rep}
    \mat U(F) = \sum_{cd} \mat M_g^T \mat U({\Frep}) \mat M_g \;,
\end{equation}
where, as in \eqnref{eq:group-acting-on-vector},
$\mat M_g^T$ is
the matrix representing
the transformation $g \in \grp$ on $\R^n$.

Finally, taking the trace of \eqnref{eq:Uab-recursive} provides a recursion
for the scalar second moment
\begin{align}\label{eq:u2ab}
    U(F_d) =
        \frac{1}{d+2}
        \sum_{F \in \children(F_d)} \! h_{\vc b}(F) \Big[
            U(F) + \|\Delta \vc b_F\|^2 \, \Vol(F)
        \Big] \;.
\end{align}

\begin{figure}\centering
    \includegraphics[width=\linewidth]{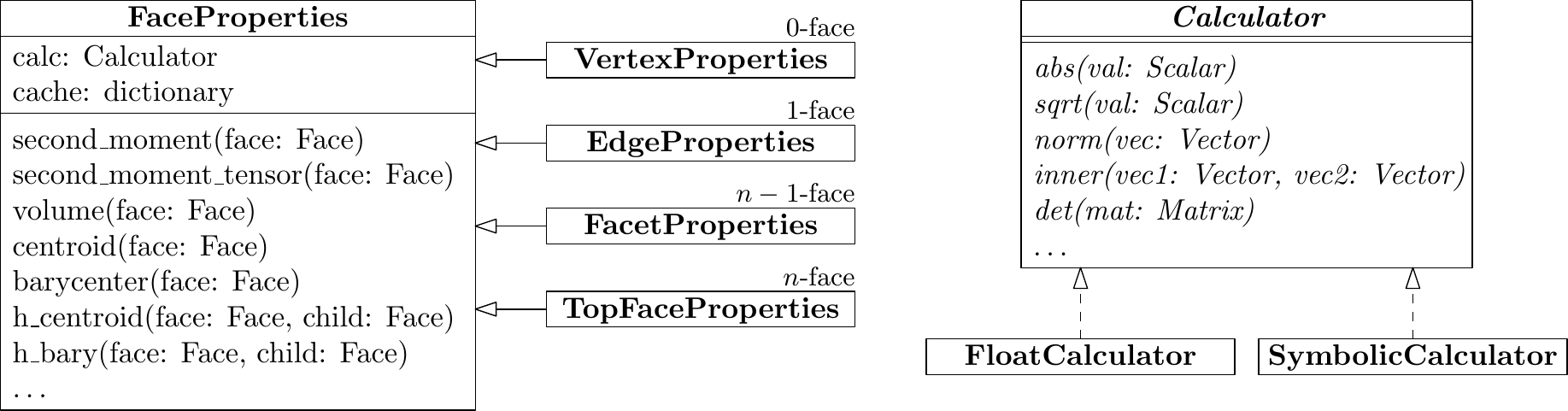}%
    \caption[]{\label{fig:property-types}%
        Classes used for implementing the calculations.
        Each \class{FaceProperties} instance stores a calculator.
        \class{FloatCalculator} and
        \class{SymbolicCalculator}
        are implementations of the abstract \class{Calculator} class
        for fast floating point and slower exact (symbolic) calculations,
        respectively.
        For certain dimensions, there are specializations of
        \class{FaceProperties} implemented as subclasses.
        See the main text for details.
    }
\end{figure}

Fig.~\ref{fig:property-types} shows the classes responsible for implementing
the recursive calculations.
Recall from Fig.~\ref{fig:data-structure} that an object of type \class{Face}
has a \field{float\_props} and a \field{sym\_props} field.
These two fields each hold a \class{FaceProperties} object,
one storing a \class{FloatCalculator} and one a
\class{SymbolicCalculator} object.

The formulas themselves are implemented in the \class{FaceProperties} class.
This class implements all calculations in terms of basic operations that are
performed by the \class{Calculator} object.
This latter object decides whether the
floating point ($\vc x^\text{float}$)
or exact symbolic ($\vc x^\text{exact}$)
components of the \class{Vector} objects are used
(see Fig.~\ref{fig:data-structure})
and performs a small set of basic operations using the correct methods for the
data type.
We use
\lib{NumPy} \cite{van_der_Walt_NumPy}
for the floating point and
\lib{SymPy} \cite{meurer2017sympy}
for the exact symbolic implementations.

Any time a property is requested from a face that is not the selected
representative of its class, the value is retrieved (and suitably
transformed) from the representative.
Such a request may happen in the recursive formulas where quantities of
child faces are needed.
Since we are guaranteed to have all children of the representative faces, the
algorithms in the \class{FaceProperties} class always have all the information
they need.

Additionally, the \class{FaceProperties} instances hold a cache of results
calculated for representative faces.
This is required to be able to reuse results of a representative for all
other faces in its class.

Certain properties have specialized algorithms for faces of dimension $0$
(vertices), $1$ (edges), $n-1$ (facets) and/or
$n$ (the top face, i.e., the Voronoi cell).
For example, a vertex has a fixed volume of $1$ and a
vanishing second moment tensor.
Similarly, the centroid of the Voronoi cell is always the origin.
The specialized methods are overridden in subclasses of the
\class{FaceProperties} class.

\section{Analyzing families of lattices}
\label{sec:results}

In this section, we apply our algorithm to calculate the second moment $G$
of an individual lattice and
to minimize $G$ in a one-parameter family of lattices.
In $9$ dimensions, such an optimization problem has led to the conjectured
optimal lattice quantizer \AEnine \cite{allen21}, which is obtained by
{\em laminating} the root lattice $D_8$.
We will first review the closely related construction of
{\em product lattices} and then the
lamination procedure.
This is then applied to optimize a family of $13$-dimensional lattices.

Product lattices are analyzed in detail in \cite{Agrell:2022jlo}.
If $\Lambda_1$ and $\Lambda_2$ are lattices of dimension $n_1$ and $n_2$,
respectively, then the one-parameter family of product lattices of
dimension $n=n_1+n_2$ is defined via
\begin{equation}\label{eq:product-lattice}
    \Lambda(a) \defeq \Lambda_1 \times a \Lambda_2 \;,
\end{equation}
where $a > 0$ is a real parameter.
Let $G(a)$ be the quantizer constant of $\Lambda(a)$ and
let further $\mat B_i$, $V_i$ and $G_i$ be, respectively, a generator matrix, the volume
and the quantizer constant of $\Lambda_i$.
Then a generator matrix for $\Lambda(a)$ is
\begin{equation}\label{eq:product-generator}
    \mat B(a) \defeq \begin{bmatrix} \mat B_1 & \mat 0 \\ \mat 0 & a \mat B_2 \end{bmatrix} .
\end{equation}
The quantizer constant of $\Lambda(a)$ is minimal when
\cite[Corollary~5]{Agrell:2022jlo}
\begin{equation}\label{eq:a-opt}
    a = \aopt \defeq \frac{V_1^{1/n_1}}{V_2^{1/n_2}} \sqrt{\frac{G_1}{G_2}} \;,
\end{equation}
for which $G(a)$ satisfies
\begin{equation}\label{eq:G-of-product}
    G^n(\aopt) = G_1^{n_1} G_2^{n_2} \;.
\end{equation}

From a collection of different lattices with known values for $G$,
Agrell and Allen form more than $30$ product lattices whose
quantizer constants are smaller than any previously published ones
in their respective dimension.
Additional improved product lattices are constructed in \cite{Lyu2022better}
in dimensions $15$, $19$, $21$ and $23$.
Starting at low dimensions $n$, the first one
found in \cite{Agrell:2022jlo} is
a product of the Coxeter--Todd lattice \Ktwelve and the one-dimensional lattice
$\Z$, i.e., $\Lambda = \KtwelveM \times a\Z$, which has
\begin{equation}\label{eq:K12-Z-G}
    G(\aopt) \approx 0.071034583 \;.
\end{equation}

The following key result motivates investigating $\KtwelveM \times a\Z$
further.
If $\Lambda_1$ and $\Lambda_2$ are two lattices with generator matrices $\mat B_1$
and $\mat B_2$ as before,
then we define $\Lambda_{\mat H}(a)$ as the lattice generated by
\begin{equation}\label{eq:gen-H}
    \mat B_{\mat H}(a) \defeq \begin{bmatrix} \mat B_1 & \mat 0 \\ \mat H & a \mat B_2 \end{bmatrix} ,
\end{equation}
where $\mat H$ is an $n_2 \times n_1$ matrix.
Let $G_{\mat H}(a)$ be the quantizer constant of $\Lambda_{\mat H}(a)$.
Then it is shown in \cite[Theorem~7]{Agrell:2022jlo} that
\begin{equation}\label{eq:G_H-less-than-G}
    G_{\mat H}(a) \leq G(a),\ \ \forall \mat H \;.
\end{equation}
This means that in general, any product lattice \eqref{eq:product-lattice}
can be further optimized by taking $\mat H \neq \mat 0$.
In particular, it should be possible to get better than \eqref{eq:K12-Z-G} in
$13$ dimensions.

The special case of the construction \eqref{eq:gen-H}
with $n_2=1$ (i.e., $\Lambda_2 = \Z$)
is called a {\em lamination} of $\Lambda_1$.
For that case, we will write
\begin{equation}\label{eq:gen-laminated}
    \mat B_{\vc h}(a) \defeq \begin{bmatrix} \mat B_1 & \mat 0 \\ \vc h & a \end{bmatrix} ,
\end{equation}
where $\vc h \in \R^{n-1}$ is the {\em offset vector}.
We will henceforth assume that $\vc h$ is fixed and drop
the subscript of the quantities $\mat B$ and $G$ of the laminated lattice
$\Lambda(a)$.

Note that for $\vc h \neq \vc 0$, we no longer have a general closed form
expression for the optimal value $\aopt$ that minimizes $G(a)$.
For example, \AEnine is obtained by laminating $D_8$ with $\vc h$ chosen as
a vertex of the Voronoi cell of $D_8$ that is most distant from $\vc 0$.
Such a vertex is called a {\em deep hole}.
It is shown in \cite{allen21} that the optimal value of $a = 0.573\ldots$ is
an algebraic number whose square is a root of a $9$th order polynomial.

In the following, we will try a similar strategy in $13$ dimensions and
laminate \Ktwelve in the direction of a deep hole.
This will require calculating $G$ as a function of $a$ and then finding the
minimum of $G(a)$.
Our method for doing so is discussed in Sec.~\ref{sub:laminatedK12}.
However, in order to verify our code and to explore the possible symmetries of
laminated \Ktwelve, we shall first apply our algorithm to \Ktwelve itself.

\subsection{The Coxeter--Todd lattice \texorpdfstring{\Ktwelve}{K12}}
\label{sub:K12}

The Coxeter--Todd lattice {\Ktwelve}
is currently
the best known lattice quantizer in $12$ dimensions
and is conjectured to be optimal \cite{conway84}.
A square generator matrix for \Ktwelve is
\cite[Section~4.9]{splag}
\begin{equation}\label{eq:K12-gen}
    \mat B \defeq \begin{bmatrix}
        2\mat A & \mat 0  & \mat 0  & \mat 0 & \mat 0 & \mat 0 \\
        \mat 0  & 2\mat A & \mat 0  & \mat 0 & \mat 0 & \mat 0 \\
        \mat 0  & \mat 0  & 2\mat A & \mat 0 & \mat 0 & \mat 0 \\
        \mat A  & \mat W  & \mat W  & \mat A & \mat 0 & \mat 0 \\
        \mat W  & \mat A  & \mat W  & \mat 0 & \mat A & \mat 0 \\
        \mat W  & \mat W  & \mat A  & \mat 0 & \mat 0 & \mat A
    \end{bmatrix} \;,
\end{equation}
where
\begin{equation}\label{eq:K12-A-W}
    \mat A \defeq \frac{1}{2} \begin{bmatrix}
        2 & 0 \\
        -1 & \qq
    \end{bmatrix} \,,\qquad
    \mat W \defeq \frac{1}{2} \begin{bmatrix}
        -1 & \qq \\
        -1 & -\qq
    \end{bmatrix}
    .
\end{equation}

\begin{centeredtable}{%
    Representatives of the relevant vectors $\vc n_i \in \normals(\Omega)$ (first two rows)
    and vertices $\vc v_i \in \verts(\Omega)$ (remaining eight rows)
    of the Voronoi cell of \Ktwelve in order of decreasing length.
    Shown are the components, squared lengths and sizes of the orbits under $\grp$.
    The vertex $\vc v_1$ has the largest squared length and thus lies
    furthest from any lattice point.
    It is used in Sec.~\ref{sub:laminatedK12} as the offset vector
    for laminating \Ktwelve.
}
    \label{tab:K12-normals-vertices}
    \begin{tabular}{l@{\qquad}r*{10}{@{\, }c@{\, }}r@{\qquad}c@{\qquad}r} \hline
        vector & \multicolumn{12}{c}{components} & $\| \cdot \|^2$ & orbit size \\
        \hline
        $\vc n_{1}$ & $\frac{1}{2}$  $(0$ & $0$ & $0$ & $0$ & $0$ & $2 \qq$ & $1$ & $\qq$ & $1$ & $\qq$ & $2$ & $0) $ & $6$ & $4\,032$ \\
        $\vc n_{2}$ &                $(0$ & $0$ & $0$ & $0$ & $0$ & $0  $ & $0$ & $0$ & $0$ & $0$ & $1$ & $-\qq)$ & $4$ & $   756$ \\
        \hline
        $\vc v_{1}$ & $\frac{1}{3} $ $(0$ & $0$ & $0$ & $0$ & $0$ & $0    $   & $0$ & $0    $   & $0 $ & $2 \qq $ & $0$ & $2 \qq)$  & $\frac{8}{3}  $ & $    20\,412$ \\
        $\vc v_{2}$ & $\frac{1}{30}$ $(0$ & $0$ & $0$ & $0$ & $3$ & $-19 \qq$ & $6$ & $-12 \qq$ & $12$ & $2 \qq $ & $9$ & $-5 \qq)$ & $\frac{52}{25}$ & $   108\,864$ \\
        $\vc v_{3}$ & $\frac{1}{15}$ $(0$ & $0$ & $0$ & $0$ & $0$ & $4 \qq  $ & $6$ & $4 \qq  $ & $12$ & $-6 \qq$ & $9$ & $-\qq)$   & $\frac{52}{25}$ & $   653\,184$ \\
        $\vc v_{4}$ & $\frac{1}{15}$ $(0$ & $0$ & $0$ & $0$ & $0$ & $4 \qq  $ & $3$ & $5 \qq  $ & $15$ & $3 \qq $ & $9$ & $\qq)$    & $\frac{52}{25}$ & $   653\,184$ \\
        $\vc v_{5}$ & $\frac{1}{9} $ $(0$ & $0$ & $0$ & $0$ & $3$ & $-5 \qq $ & $3$ & $\qq    $ & $3 $ & $-3 \qq$ & $3$ & $3 \qq)$  & $\frac{56}{27}$ & $   326\,592$ \\
        $\vc v_{6}$ & $\frac{1}{9} $ $(0$ & $0$ & $0$ & $0$ & $0$ & $2 \qq  $ & $3$ & $3 \qq  $ & $6 $ & $-4 \qq$ & $6$ & $0)$      & $\frac{56}{27}$ & $3\,265\,920$ \\
        $\vc v_{7}$ & $\frac{1}{9} $ $(0$ & $0$ & $0$ & $0$ & $0$ & $2 \qq  $ & $3$ & $-\qq   $ & $3 $ & $3 \qq $ & $9$ & $3 \qq)$  & $\frac{56}{27}$ & $   653\,184$ \\
        $\vc v_{8}$ & $\frac{1}{6} $ $(0$ & $0$ & $0$ & $0$ & $0$ & $2 \qq  $ & $3$ & $-\qq   $ & $3 $ & $-\qq  $ & $6$ & $0)$      & $2            $ & $     4\,032$ \\
        \hline
    \end{tabular}
\end{centeredtable}

Its symmetry group is described in \cite[Section~4.9]{splag}
and \cite{conway1983coxeter}
and it has order $|\grp| = 78\,382\,080$.
These references also discuss the $4\,788$ relevant vectors and the
deep holes.
Tab.~\ref{tab:K12-normals-vertices} lists the two representatives
$\vc n_1$, $\vc n_2$ of the
relevant vectors and the single representative deep hole $\vc v_1$
along with representatives $\vc v_2, \ldots, \vc v_8$ of all other vertices of the
Voronoi cell of \Ktwelve.
We obtained the full set of $5\,685\,372$
vertices using the method discussed in Sec.~\ref{sub:vertices}.

With the help of \lib{GAP}, we determined three matrices that together
generate the full group $\grp$, namely
\begin{equation}\label{eq:K12-group-gens}
    \begin{aligned}[b]
        \mat M_1 &\defeq \begin{bmatrix}
            \mat 0   & \mat I_2 & \mat 0   & \mat 0   & \mat 0   & \mat 0 \\
            \mat I_2 & \mat 0   & \mat 0   & \mat 0   & \mat 0   & \mat 0 \\
            \mat 0   & \mat 0   & \mat I_2 & \mat 0   & \mat 0   & \mat 0 \\
            \mat 0   & \mat 0   & \mat 0   & \mat 0   & \mat I_2 & \mat 0 \\
            \mat 0   & \mat 0   & \mat 0   & \mat I_2 & \mat 0   & \mat 0 \\
            \mat 0   & \mat 0   & \mat 0   & \mat 0   & \mat 0   & \mat I_2
        \end{bmatrix} ,\quad&
        \mat M_2 &\defeq \begin{bmatrix}
            \mat 0 & \mat 0 & \mat 0 & \mat 0 & \mat 0 & \mat S \\
            \mat 0 & \mat 0 & \mat 0 & \mat 0 & \mat S & \mat 0 \\
            \mat 0 & \mat 0 & \mat 0 & \mat S & \mat 0 & \mat 0 \\
            \mat 0 & \mat 0 & \mat S & \mat 0 & \mat 0 & \mat 0 \\
            \mat 0 & \mat S & \mat 0 & \mat 0 & \mat 0 & \mat 0 \\
            \mat S & \mat 0 & \mat 0 & \mat 0 & \mat 0 & \mat 0
        \end{bmatrix} ,\\
        \mat M_3 &\defeq \frac{1}{2} \begin{bmatrix}
            \mat I_2  & \mat V    & -\mat I_2 & \mat 0    & \mat V    & \mat 0 \\
            \mat V^T  & \mat I_2  & \mat V^T  & \mat 0    & -\mat I_2 & \mat 0 \\
            -\mat I_2 & \mat V    & \mat I_2  & \mat 0    & \mat V    & \mat 0 \\
            \mat 0    & \mat 0    & \mat 0    & 2\mat I_2 & \mat 0    & \mat 0 \\
            \mat V^T  & -\mat I_2 & \mat V^T  & \mat 0    & \mat I_2  & \mat 0 \\
            \mat 0    & \mat 0    & \mat 0    & \mat 0    & \mat 0    & 2\mat I_2
        \end{bmatrix} ,
    \end{aligned}
\end{equation}
where $\mat I_k$ is the $k \times k$ identity matrix and
\begin{equation}\label{eq:K12-group-gen-sub}
    \begin{aligned}[b]
        \mat S &\defeq \begin{bmatrix} 1 & 0 \\ 0 & -1 \end{bmatrix} , \qquad&
        \mat V &\defeq \frac{1}{2} \begin{bmatrix}
            1 & \qq \\ -\qq & 1
        \end{bmatrix}.
    \end{aligned}
\end{equation}

The construction of the hierarchy of faces is done as described in
Sec.~\ref{sec:voronoi-construction}.
We employ the iterated classification using proper subgroups
$\setU_i \subset \grp$ followed by the full group $\grp$.
One could easily obtain such subgroups by taking any proper subset of the
three generator matrices $\{\mat M_1, \mat M_2, \mat M_3\}$ as generators of a subgroup.
However, the subgroups generated by a single of these matrices
each have order $2$,
while the subgroups generated by any pair of matrices have
orders $8$ ($\mat M_1$ and $\mat M_2$) or $12$ (the other pairs).
These subgroups are too small to efficiently reduce the number of faces by
consolidating them into classes.

Larger subgroups $\setU_i \subset \grp$ are, e.g., the stabilizers of
relevant vectors---and thus of the facets---or more generally of any subface $F_d$.
We chose the stabilizers of the two representative facets, i.e., of $\vc n_1$
and $\vc n_2$ shown in Tab.~\ref{tab:K12-normals-vertices},
with sizes $19\,440$ and $103\,680$, respectively.
Using these two subgroups and $\grp$,
we classify the children of the representative facets, yielding
six classes of $10$-faces.
The stabilizers of these six representatives have sizes
$240$, $480$, $648$, $1\,296$, $2\,592$ and $103\,680$,
where we note that the two subgroups of size $103\,680$ are indeed distinct.
We now have eight subgroups $\setU_i$ and the full group $\grp$
with which to perform the iterated
classification in all lower dimensions.

The resulting face hierarchy contains $809$ classes of faces.
From dimension $1$ through $12$, the number of classes is
$8$, $22$, $48$, $93$, $149$, $185$, $154$, $86$, $40$, $15$, $6$, $2$ and $1$,
respectively.

We calculate an exact value for the quantizer constant of \Ktwelve,
\begin{equation}\label{eq:K12-G}
    G = \frac{797\,361\,941}{6\,567\,561\,000 \, \sqrt{3}}
        \approx 0.0700956
        \;,
\end{equation}
which agrees with the value reported in \cite{DSSV09}.
We also calculate the (unnormalized) second moment tensor
\begin{equation}\label{eq:K12-Uab}
    \mat U = \frac{797\,361\,941}{243\,243\,000} \mat I_{12}
        \approx 3.2780468 \, \mat I_{12} \;.
\end{equation}
Recall that due to \cite{Agrell:2022jlo}, a locally optimal lattice quantizer
necessarily has a second moment tensor proportional to the identity matrix.
Despite \Ktwelve being the best currently known lattice quantizer in $12$
dimensions, it is
not known if it is globally or at least locally optimal.
The result \eqref{eq:K12-Uab} therefore supports that \Ktwelve may be
at least locally optimal.

A comprehensive catalog of all $809$ face classes with exact expressions for
volumes, second moment scalars and hierarchical information is available as
a supplementary online resource \cite{ancillary-files}.

\subsection{Laminated \texorpdfstring{\Ktwelve}{K12}}
\label{sub:laminatedK12}

In this section, we apply our algorithm to a lamination $\Lambda(a)$ of \Ktwelve.
We take as offset vector $\vc h$
a deep hole of \Ktwelve,
$\vc v_1$
(see Tab.~\ref{tab:K12-normals-vertices}), i.e., we have the generator matrix
\begin{equation}\label{eq:lK12-gen}
    \mat B(a) \defeq \begin{bmatrix}
        \mat B_1 & \mat 0 \\ \vc v_1 & a
    \end{bmatrix} ,
\end{equation}
where $\mat B_1$ is the $12 \times 12$ generator matrix \eqref{eq:K12-gen} of
\Ktwelve.

Note that it is unclear whether our choice of $\vc h = \vc v_1$ is optimal.
However, as we shall see below, with an optimal $a$
this choice leads to a second moment tensor
$\mat U$ that is proportional to the identity matrix,
supporting that $\Lambda(a)$ might at least be a
locally optimal lattice quantizer.

Since we are interested in finding the minimum of $G$, we begin our analysis by
numerically estimating $G(a)$ via Monte Carlo integration for several values
of $a$.
A resulting rough estimate $a_0 \approx \aopt$ is then used
for constructing the Voronoi cell (Sec.~\ref{sec:voronoi-construction})
and performing the calculations (Sec.~\ref{sec:second-moment}).
In the present case of laminated \Ktwelve, we used $a_0 = 34/33 \approx 1.03$.

We next carry out the full analysis at $a_0$
and determine $G$ as well as the volumes $\Vol(F_d)$ of all faces $F_d$ as
functions of $a$.
This can be done via symbolic calculations of the equations in
Sec.~\ref{sec:second-moment}.
We remark that instead of symbolically calculating with an unknown $a$,
one may speed up the calculations substantially by substituting an (exact)
rational value for $a$.
Following the discussion in \cite[Sec.~6]{allen21}, we used
$n+3$ rational values close to $a_0$ to initially determine $G(a)$ and
later verified our results with full symbolic calculation with
unknown $a$.

At this point, we do not yet know for which values of $a$ these results hold.
As $a$ changes, vertices may merge or split,
resulting in changes to the hierarchy of faces.
At these {\em critical} values of $a$, not only may the functional dependence
of $G$ on $a$ change (phase transition),
the whole data structure we build for one value of $a$ becomes invalid.

We perform the following steps to determine the domain of $a$ where our
results apply.
This is done after the data structure has been calculated at
$a=a_0$ and the volumes of all $d$-faces, $d = 0, \ldots, n$, have been determined
as function of $a$.
First,
we find an interval $I=(a_-,a_+)$ around $a_0$ where
the expressions for all the volumes evaluate to strictly positive values.
Our data structure evaluated at any $a \in I$ then represents a convex polytope
$P(a)$ with faces having positive volumes.
In particular, the volume $\Vol(P(a))$ is correctly calculated by
\eqnref{eq:volume-recursive}.
We have $P(a_0) = \Omega(a_0)$, where $\Omega(a)$ is the Voronoi cell of $\Lambda(a)$,
but away from $a_0$, $P(a)$ does not need to agree with $\Omega(a)$.
In principle, vertices of $\Omega(a)$ can split in this interval and new faces may appear,
so that the vertices of $P(a)$ need not coincide with those of
$\Omega(a)$ for all
$a \in I$.
Therefore, we additionally check the following
two conditions: ($i$)
The set of relevant vectors found at $a_0$
is the (full) set of relevant vectors of $\Omega(a)$ for any $a \in I$.
($ii$) The
representative vertices are still inside
$\Omega(a)$ and contained in the same facets as at $a=a_0$.
This is verified using \eqnref{eq:vertex-conditions} and
\eqnref{eq:vertex-in-facet-float}.
Then, $P(a) \subseteq \Omega(a)$
and if furthermore $\Vol(P(a)) = \lvert\det \mat B\rvert$,
then clearly $P(a) = \Omega(a)$.

For laminated \Ktwelve analyzed at $a_0 = 34/33 \approx 1.03$,
we find vanishing volumes of $1$-faces of $P(a)$ at $a = 1$ and
$a = \sqrt{17/15} \approx 1.06$.
Faces have non-vanishing volumes between these two values and the calculated
volume of $\Omega$ is always $\lvert\det \mat B\rvert = 27a$.
Also, the additional conditions ($i$) and ($ii$)
mentioned above are satisfied in that range.
This means that the formulas we report below are valid for
\begin{equation}\label{eq:lK12-a-range}
    1 \leq a \leq \sqrt{17/15}\;.
\end{equation}

The analysis itself starts by finding all the relevant vectors, which is
step~1 of Sec.~\ref{sub:naive}.
Using the algorithm presented in \cite{AEVZ}, we obtain a set of $7\,706$
vectors $\normals(\Omega)$.
At this point, they are not yet classified, since we do not know the
symmetry group $\grp$ of laminated \Ktwelve.

However, the relevant vectors themselves help in finding $\grp$ as follows.
We take the symmetry group $\grp_{12}$ of \Ktwelve and
embed it in $O(13)$ via
\begin{equation}\label{eq:lK12-embed-grp}
    \grp_{12 \hookrightarrow 13} \defeq \left\{
            \begin{bmatrix} \mat M_g^T & \mat 0 \\ \mat 0 & 1 \end{bmatrix} : g \in \grp_{12}
        \right\} \;,
\end{equation}
where, as before, $\mat M_g^T$ is the $12 \times 12$ matrix representing $g \in \grp_{12}$.
This group is generated by the matrices
\begin{equation}\label{eq:gens-of-G1213}
    \begin{aligned}[b]
        \begin{bmatrix}
            \mat M^{(12)}_1 & \mat 0 \\ \mat 0 & 1
        \end{bmatrix} , \quad&&
        \begin{bmatrix}
            \mat M^{(12)}_2 & \mat 0 \\ \mat 0 & 1
        \end{bmatrix} , \quad&&
        \begin{bmatrix}
            \mat M^{(12)}_3 & \mat 0 \\ \mat 0 & 1
        \end{bmatrix} ,
    \end{aligned}
\end{equation}
where $\mat M^{(12)}_i$ are the generators \eqref{eq:K12-group-gens} of \Ktwelve.
We add a fourth generator
\begin{equation}\label{eq:add-gen-to-G1213}
    \begin{bmatrix}
        \mat I_{12} & \mat 0 \\ \mat 0 & -1
    \end{bmatrix}
\end{equation}
to create a group $\bar\grp$ of twice the order of $\grp_{12}$.
Then, we use \lib{GAP} to compute the stabilizer
\begin{equation}\label{eq:lK12-group-stab}
    \grp \defeq \left\{
        g \in \bar\grp
            : g \normals(\Omega) = \normals(\Omega)
    \right\}
\end{equation}
of the set of relevant vectors of $\Omega$ and produce a generating set for
$\grp$.
This way, we obtain the three symmetries
\begin{align}
    \label{eq:K12-group-M1}
    \mat M_1 &\defeq \frac{1}{2} \begin{bmatrix}
        \mat 0    & \mat I_2 & \mat 0    & \mat V    & -\mat I_2 & -\mat V  & \mat 0 \\
        \mat 0    & -\mat V  & \mat V    & \mat 0    & -\mat I_2 & \mat I_2 & \mat 0 \\
        2\mat V^T & \mat 0   & \mat 0    & \mat 0    & \mat 0    & \mat 0   & \mat 0 \\
        \mat 0    & \mat 0   & -\mat I_2 & \mat V    & \mat V    & \mat I_2 & \mat 0 \\
        \mat 0    & \mat I_2 & \mat V    & -\mat I_2 & \mat V    & \mat 0   & \mat 0 \\
        \mat 0    & \mat V   & \mat I_2  & \mat I_2  & \mat 0    & \mat V   & \mat 0 \\
        \mat 0    & \mat 0   & \mat 0    & \mat 0    & \mat 0    & \mat 0   & 2
    \end{bmatrix} , \\
    \label{eq:K12-group-M2}
    \mat M_2 &\defeq \frac{1}{2} \begin{bmatrix}
        \mat 0  & -\mat S & \mat 0   & \mat Y  & -\mat S  & \mat Y  & \mat 0 \\
        2\mat S & \mat 0  & \mat 0   & \mat 0  & \mat 0   & \mat 0  & \mat 0 \\
        \mat 0  & \mat Y' & \mat 0   & -\mat S & -\mat Y' & \mat S  & \mat 0 \\
        \mat 0  & \mat 0  & -2\mat S & \mat 0  & \mat 0   & \mat 0  & \mat 0 \\
        \mat 0  & \mat Y' & \mat 0   & \mat S  & -\mat Y' & -\mat S & \mat 0 \\
        \mat 0  & -\mat S & \mat 0   & -\mat Y & -\mat S  & -\mat Y & \mat 0 \\
        \mat 0  & \mat 0  & \mat 0   & \mat 0  & \mat 0   & \mat 0  & 2
    \end{bmatrix} , \\
    \label{eq:K12-group-M3}
    \mat M_3 &\defeq \begin{bmatrix} \mat I_8 & \mat 0 \\ \mat 0 & -\mat I_5 \end{bmatrix} ,
\end{align}
where $\mat S$ and $\mat V$ are given in \eqnref{eq:K12-group-gen-sub}
and
\begin{equation}\label{eq:lK12-gen-sub}
    \begin{aligned}[b]
        \mat Y &\defeq \frac{1}{2} \begin{bmatrix} -1 & -\qq \\ -\qq & 1 \end{bmatrix} ,\qquad&
        \mat Y' &\defeq \frac{1}{2} \begin{bmatrix} -1 & \qq \\ \qq & 1 \end{bmatrix} .
    \end{aligned}
\end{equation}
This group $\grp$, generated by
$\mat M_1$, $\mat M_2$ and $\mat M_3$, has order $622\,080$
and we use it as the symmetry group of laminated \Ktwelve for the next steps.
In principle, more symmetries might exist, in particular for specific values of $a$,
resulting in a larger symmetry group.
However, we were not able to find any for generic $a$.

\begin{centeredtable}{%
    Representatives of the $7\,706$ relevant vectors $\vc n_i \in \normals(\Omega)$
    of the Voronoi cell of laminated \Ktwelve.
    The columns are as for Tab.~\ref{tab:K12-normals-vertices}.
    The $482$ representatives of the vertices of $\Omega$ are not shown here.
}
    \label{tab:LK12-normals}
    \begin{tabular}{l@{ }r*{11}{@{ }c@{ }}r@{\quad}c@{\quad}r} \hline
        vector & \multicolumn{13}{c}{components} & $\|\vc n_i\|^2$ & orbit size \\
        \hline
        $\vc n_{1}$ & $            (0$ & $0$ & $0$ & $0$ & $0$ & $0$ & $0$ & $0$ & $0$ & $0$ & $0$ & $0$ & $- 3 a)$ &$9 a^{2}$ & $2$ \\
        $\vc n_{2}$ & $\frac{1}{3} (0$ & $0$ & $0$ & $0$ & $0$ & $0$ & $0$ & $0$ & $0$ & $- 2 \qq$ & $0$ & $- 2 \qq$ & $6 a)$ &$4 a^{2} + \frac{8}{3}$ & $162$ \\
        $\vc n_{3}$ & $\frac{1}{2} (-3$ & $- \qq$ & $-2$ & $0$ & $0$ & $0$ & $-2$ & $0$ & $0$ & $0$ & $-1$ & $\qq$ & $0)$ &$6$ & $2\,592$ \\
        $\vc n_{4}$ & $\frac{1}{2} (-3$ & $- \qq$ & $0$ & $0$ & $1$ & $\qq$ & $0$ & $0$ & $-1$ & $- \qq$ & $1$ & $- \qq$ & $0)$ &$6$ & $1\,440$ \\
        $\vc n_{5}$ & $\frac{1}{6} (-3$ & $- 3 \qq$ & $-3$ & $3 \qq$ & $-3$ & $3 \qq$ & $-3$ & $- 3 \qq$ & $0$ & $2 \qq$ & $0$ & $2 \qq$ & $- 6 a)$ &$a^{2} + \frac{14}{3}$ & $2\,592$ \\
        $\vc n_{6}$ & $\frac{1}{2} (-2$ & $0$ & $-1$ & $\qq$ & $-1$ & $\qq$ & $-2$ & $0$ & $0$ & $0$ & $0$ & $0$ & $0)$ &$4$ & $216$ \\
        $\vc n_{7}$ & $\frac{1}{2} (-2$ & $0$ & $-2$ & $0$ & $1$ & $\qq$ & $0$ & $0$ & $0$ & $0$ & $-1$ & $- \qq$ & $0)$ &$4$ & $540$ \\
        $\vc n_{8}$ & $\frac{1}{6} (0$ & $0$ & $3$ & $- 3 \qq$ & $0$ & $0$ & $6$ & $0$ & $-3$ & $- \qq$ & $0$ & $2 \qq$ & $- 6 a)$ &$a^{2} + \frac{8}{3}$ & $162$ \\
        \hline
    \end{tabular}
\end{centeredtable}

Using $\grp$, the $7\,706$ relevant vectors are partitioned into $8$ classes,
representatives of which are listed in Tab.~\ref{tab:LK12-normals}.
The search for vertices of $\Omega$ (step 2) is done as described in
Sec.~\ref{sub:vertices} and yields
$52\,351\,632$ vertices in $482$ classes.

As for \Ktwelve, we employ the iterated classification method and therefore
need subgroups $\setU_i$ of $\grp$.
The subgroups generated by only $\mat M_1$, $\mat M_2$ and $\mat M_3$ have orders
$30$, $10$ and $2$, respectively.
Those generated by the pairs
$\{\mat M_1, \mat M_2\}$
and
$\{\mat M_1, \mat M_3\}$ both have order $311\,040$, while the pair
$\{\mat M_2, \mat M_3\}$ generates a group of order $320$.
Again, instead of these, we use the stabilizers of the representative facets
and $(n-2)$-faces as subgroups $\setU_i$.
The $8$ facets (see Tab.~\ref{tab:LK12-normals}) have stabilizer sizes ranging
from $240$ to $3\,840$ and the $81$ faces of dimension $11$ have stabilizers
with sizes between $4$ and $3\,840$.
Of these in total $89$ subgroups, we selected $13$ with sizes
$12$,
$48$,
$480$,
$1\,152$,
$1\,920$,
$2\,880$ (twice) and
$3\,840$ (six times)
and used these for the iterated classification.
We remark that the decision on how to construct and choose the subgroups can
likely be optimized further.

Our construction of the hierarchy of faces
resulted in $430\,051$ classes.
From dimensions $0$ to $13$, the number of classes is
$482$,
$3\,599$,
$15\,656$,
$45\,473$,
$87\,511$,
$110\,578$,
$92\,074$,
$50\,820$,
$18\,590$,
$4\,477$,
$701$,
$81$,
$8$ and
$1$.

The total number of face classes of laminated \Ktwelve is much larger
than for \Ktwelve (which has $809$ face classes)
in part due to the much smaller symmetry group.
If the group were larger, more faces would potentially be
equivalent.
On the other hand,
the smaller symmetry group also leads to more but smaller
subsets $\XX_j$
of the defining sets $\defD(F_d)$ of vectors we use for the classification
(see Sec.~\ref{sub:defining-vectors} and \ref{sub:find-transform}).
This reduces the classification cost for an individual face.
However, the much larger number of classes in the end leads to a much more
expensive construction of the face hierarchy than for \Ktwelve.

For a full catalog of face classes, the reader is again referred to the
supplementary online material \cite{ancillary-files}, where we list properties and hierarchical information
of all face classes.

Using the formulas of Sec.~\ref{sec:second-moment},
the unnormalized second moment $U$ as a function of $a$
is
\begin{gather}\label{eq:lK12-U}
    \begin{aligned}[b]
        U ={}
        & \tfrac{1239510953 a^{27}}{58118860800} - \tfrac{4417638557 a^{25}}{12454041600}
            + \tfrac{15554872313 a^{23}}{5748019200} - \tfrac{54014687957 a^{21}}{4311014400}\\
        &+ \tfrac{184704081953 a^{19}}{4702924800} - \tfrac{621669488957 a^{17}}{7054387200}
            + \tfrac{2064191975273 a^{15}}{14108774400} - \tfrac{6819548630117 a^{13}}{37035532800}\\
        &+ \tfrac{22891056666353 a^{11}}{126978969600} - \tfrac{81231044680397 a^{9}}{571405363200}
            + \tfrac{321224675816633 a^{7}}{3428432179200} - \tfrac{1463538531346037 a^{5}}{28284565478400}\\
        &+ \tfrac{8313434653636289 a^{3}}{339414785740800} + \tfrac{218456407528702627 a}{6618588321945600}
            + \tfrac{9696717442377617}{884484075750912000 a}
        \;.
    \end{aligned}
\end{gather}

With the volume $\Vol(\Lambda) = 27a$ and using $E=U/\Vol(\Lambda)$ it is easy to calculate $G(a)$ via
\eqnref{eq:G}.
The condition $G'(a) = 0$
can be turned into a polynomial equation
$f(v) = 0$, where $v \defeq a^2$ and
\begin{equation}\label{eq:lK12-a_opt-poly}
    \begin{aligned}[b]
        f(v) \defeq{}
        &\tfrac{1239510953 v^{14}}{4151347200} - \tfrac{136946795267 v^{13}}{29889699840}
            + \tfrac{1104395934223 v^{12}}{34488115200} - \tfrac{2322631582151 v^{11}}{17244057600}\\
        &+ \tfrac{5356418376637 v^{10}}{14108774400} - \tfrac{64031957362571 v^{9}}{84652646400}
            + \tfrac{2064191975273 v^{8}}{1881169920} - \tfrac{75015034931287 v^{7}}{63489484800}\\
        &+ \tfrac{22891056666353 v^{6}}{23808556800} - \tfrac{1380927759566749 v^{5}}{2285621452800}
            + \tfrac{6103268840516027 v^{4}}{20570593075200} - \tfrac{1463538531346037 v^{3}}{13576591429632}\\
        &+ \tfrac{8313434653636289 v^{2}}{339414785740800} - \tfrac{218456407528702627 v}{79423059863347200}
            - \tfrac{9696717442377617}{758129207786496000}
        \;.
    \end{aligned}
\end{equation}
If $v_0$ denotes the smallest positive root of $f$, then the minimum of $G(a)$ is
attained at
\begin{equation}\label{eq:lK12-aopt}
    \aopt = \sqrt{v_0} \approx 1.0149980107 \;,
\end{equation}
which lies in the range \eqref{eq:lK12-a-range}.
The resulting value of the quantizer constant is
\begin{equation}\label{eq:lK12-Gopt}
    G(\aopt) \approx 0.0699012856 \;.
\end{equation}
This lies well below the currently known best second moment \eqref{eq:K12-Z-G}
in $13$ dimensions
\cite[Table~I]{Agrell:2022jlo}, \cite[Table~I]{Lyu2022better}.

The second moment tensor is the diagonal $13 \times 13$
matrix
\begin{equation}
    \mat U = \alpha(a) \mat I_{13} + \beta(a) \mat Z_{13}
    \;,
\end{equation}
where $\mat Z_{13} = \Diag(0,\ldots,0,1)$,
\begin{gather}\label{eq:lK12-Uab-alpha}
    \begin{aligned}[b]
        \alpha(a) ={}
        &- \tfrac{1239510953 a^{27}}{58118860800} + \tfrac{4417638557 a^{25}}{13586227200}
            - \tfrac{15554872313 a^{23}}{6897623040} + \tfrac{54014687957 a^{21}}{5748019200}\\
        &- \tfrac{184704081953 a^{19}}{7054387200} + \tfrac{621669488957 a^{17}}{12093235200}
            - \tfrac{2064191975273 a^{15}}{28217548800} + \tfrac{6819548630117 a^{13}}{88885278720}\\
        &- \tfrac{22891056666353 a^{11}}{380936908800} + \tfrac{81231044680397 a^{9}}{2285621452800}
            - \tfrac{321224675816633 a^{7}}{20570593075200} + \tfrac{1463538531346037 a^{5}}{339414785740800}\\
        &+ \tfrac{218456407528702627 a}{79423059863347200} + \tfrac{9696717442377617}{5306904454505472000 a}
    \end{aligned}
\end{gather}
and
\begin{gather}\label{eq:lK12-Uab-beta}
    \begin{aligned}[b]
        \beta(a) ={}
        \frac{f(a^2)}{a}
        \;.
    \end{aligned}
\end{gather}
As was done for $G(a)$ initially, we obtained \eqnref{eq:lK12-Uab-alpha} and
\eqnref{eq:lK12-Uab-beta} by following \cite[Sec.~6]{allen21} to infer
$\mat U$ as
a function of $a$ from $n+3$ exact results for rational $a$.

Note that
\begin{equation}\label{eq:beta-propto-Uprime}
    \beta(a) \propto a^\frac{28}{13} G'(a)
    \;,
\end{equation}
which shows that
$\mat U$
is proportional to the identity matrix
if and only if $G'(a) = 0$.
This is satisfied at $a = \aopt$.

\section{Conclusions}
\label{sec:conclusions}

For a lattice whose symmetry group is known,
the algorithm presented in this work provides a way to explicitly construct a
part of the full face hierarchy of its Voronoi cell.
This partial structure contains sufficient information for exactly
calculating, among other properties, its quantizer constant $G$.
In addition, one obtains a full classification of all faces of the cell,
despite most faces never being constructed in this process.

A key property of our approach is the possibility to parallelize the
algorithm,
which enables the work to be distributed over many
processor cores.
By carefully planning the steps needed to evaluate the recursion formulas
for $G$, parallelization can be employed not only during
construction of the face hierarchy, but also in the subsequent
computations.

We have applied our algorithm to reproduce the results known for \AEnine and
\Ktwelve and then analyzed a new $13$-dimensional family of lattices obtained
by laminating \Ktwelve.
This lead to a new currently best known lattice quantizer in $13$ dimensions.
With $430\,051$ different classes of faces, analyzing this family with our
method proved to still be computationally feasible.
The most expensive part is determining
whether two faces are equivalent under the lattice's symmetry group and
obtaining a group element transforming one into the other.
If a more efficient method can be found for this step, then most of the
strategy discussed in this paper is still applicable
and may enable analysis of lattices in significantly higher dimensions.

\bibliographystyle{siamplain}
\bibliography{references}

\begin{thebibliography}{10}

\bibitem{Agrell:2022jlo}
{\sc E.~Agrell and B.~Allen}, {\em {On the best lattice quantizers}}, February
  2022, \url{https://arxiv.org/abs/2202.09605}.

\bibitem{agrell98}
{\sc E.~Agrell and T.~Eriksson}, {\em Optimization of lattices for
  quantization}, {IEEE} Trans. Inf. Theory, 44 (1998), pp.~1814--1828,
  \url{https://doi.org/10.1109/18.705561}.

\bibitem{AEVZ}
{\sc E.~{Agrell}, T.~{Eriksson}, A.~{Vardy}, and K.~{Zeger}}, {\em Closest
  point search in lattices}, {IEEE} Trans. Inf. Theory, 48 (2002),
  pp.~2201--2214, \url{https://doi.org/10.1109/TIT.2002.800499}.

\bibitem{allen2022performance}
{\sc B.~Allen}, {\em Performance of random template banks}, Phys. Rev. D, 105
  (2022), p.~102003, \url{https://doi.org/10.1103/PhysRevD.105.102003}.

\bibitem{allen21}
{\sc B.~Allen and E.~Agrell}, {\em The optimal lattice quantizer in nine
  dimensions}, Annalen der Physik, 533 (2021), p.~2100259,
  \url{https://doi.org/10.1002/andp.202100259}.

\bibitem{ancillary-files}
The arXiv version of this paper has ancillary files containing catalogs of the
  face classes of $K_{12}$ and laminated $K_{12}$.

\bibitem{barnes83}
{\sc E.~S. Barnes and N.~J.~A. Sloane}, {\em The optimal lattice quantizer in
  three dimensions}, SIAM J. Alg. Disc. Meth., 4 (1983), pp.~30--41,
  \url{https://doi.org/10.1137/0604005}.

\bibitem{gappySoftware}
{\sc E.~M. Bray}, {\em gappy {\textendash} a {P}ython interface to {GAP},
  {V}ersion 0.1.0a4}, \url{https://github.com/embray/gappy}.

\bibitem{conway1983coxeter}
{\sc J.~H. Conway and N.~J.~A. Sloane}, {\em The {C}oxeter--{T}odd lattice, the
  {M}itchell group, and related sphere packings}, in Mathematical Proceedings
  of the Cambridge Philosophical Society, vol.~93, Cambridge University Press,
  1983, pp.~421--440, \url{https://doi.org/10.1017/S0305004100060746}.

\bibitem{conway84}
{\sc J.~H. Conway and N.~J.~A. Sloane}, {\em On the {Voronoi} regions of
  certain lattices}, SIAM J. Alg. Disc. Meth., 5 (1984), pp.~294--305,
  \url{https://doi.org/10.1137/0605031}.

\bibitem{conway85}
{\sc J.~H. Conway and N.~J.~A. Sloane}, {\em A lower bound on the average error
  of vector quantizers}, {IEEE} Trans. Inf. Theory, IT-31 (1985), pp.~106--109,
  \url{https://doi.org/10.1109/TIT.1985.1056993}.

\bibitem{splag}
{\sc J.~H. Conway and N.~J.~A. Sloane}, {\em Sphere Packings, Lattices and
  Groups}, Springer, New York, NY, 3rd~ed., 1999,
  \url{https://doi.org/10.1007/978-1-4757-6568-7}.

\bibitem{gersho79}
{\sc A.~Gersho}, {\em Asymptotically optimal block quantization}, {IEEE} Trans.
  Inf. Theory, IT-25 (1979), pp.~373--380,
  \url{https://doi.org/10.1109/TIT.1979.1056067}.

\bibitem{Lyu2022better}
{\sc S.~Lyu, Z.~Wang, C.~Ling, and H.~Chen}, {\em Better lattice quantizers
  constructed from complex integers}, April 2022,
  \url{https://arxiv.org/abs/2204.01105}.

\bibitem{martinet13book}
{\sc J.~Martinet}, {\em Perfect lattices in Euclidean spaces}, vol.~327,
  Springer Science \& Business Media, 2013.

\bibitem{meurer2017sympy}
{\sc A.~Meurer et~al.}, {\em {SymPy}: symbolic computing in {Python}}, PeerJ
  Computer Science, 3 (2017), p.~e103,
  \url{https://doi.org/10.7717/peerj-cs.103}.

\bibitem{minkowski1989allgemeine}
{\sc H.~Minkowski}, {\em {Allgemeine Lehrs{\"a}tze {\"u}ber die convexen
  Polyeder}}, Nachrichten von der K{\"o}nigl.\ Gesellschaft der Wissenschaften
  zu G{\"o}ttingen. Mathematisch-physikalische Klasse,  (1897), pp.~198--219,
  \url{http://resolver.sub.uni-goettingen.de/purl?PPN252457811_1897}.

\bibitem{DSSV09}
{\sc M.~D. Sikirić, A.~Schürmann, and F.~Vallentin}, {\em {Complexity and
  algorithms for computing Voronoi cells of lattices}}, Mathematics of
  Computation, 78 (2009), p.~1713–1731,
  \url{https://doi.org/10.1090/s0025-5718-09-02224-8}.

\bibitem{GAPSoftware}
{\sc {The GAP {G}roup}}, {\em {GAP} {\textendash} {G}roups, {A}lgorithms, and
  {P}rogramming, {V}ersion 4.12dev}, \url{https://www.gap-system.org}.

\bibitem{GAPRepo}
{\sc {The GAP {G}roup}}, {\em {Main development repository for GAP
  {\textendash} Groups, Algorithms, Programming}}, October 2021,
  \url{https://github.com/gap-system/gap}.
\newblock {Commit 401c797476b787e748a3890be4ce95ae4e5d52ae}.

\bibitem{torquato2010reformulation}
{\sc S.~Torquato}, {\em Reformulation of the covering and quantizer problems as
  ground states of interacting particles}, Phys. Rev. E, 82 (2010), p.~056109,
  \url{https://doi.org/10.1103/PhysRevE.82.056109}.

\bibitem{van_der_Walt_NumPy}
{\sc S.~{van der Walt}, S.~C. {Colbert}, and G.~{Varoquaux}}, {\em The {NumPy}
  array: A structure for efficient numerical computation}, Computing in Science
  Engineering, 13 (2011), pp.~22--30,
  \url{https://doi.org/10.1109/MCSE.2011.37}.

\bibitem{ViterboBiglieri}
{\sc E.~Viterbo and E.~Biglieri}, {\em Computing the {Voronoi} cell of a
  lattice: The diamond-cutting algorithm}, {IEEE} Trans. Inf. Theory, 42
  (1996), pp.~161--171, \url{https://doi.org/10.1109/18.481786}.

\bibitem{Voronoi4}
{\sc G.~Vorono\"i}, {\em {Nouvelles applications des param\`etres continus \`a
  la th\'eorie des formes quadratiques, Deuxi\`eme M\'emoire, Recherches sur
  les parall\'elo\`edres primitifs}}, {J. Reine Angew. Math.}, 136 (1909),
  pp.~67--181, \url{https://doi.org/doi:10.1515/crll.1909.136.67}.

\bibitem{zador82}
{\sc P.~L. Zador}, {\em Asymptotic quantization error of continuous signals and
  the quantization dimension}, {IEEE} Trans. Inf. Theory, IT-28 (1982),
  pp.~139--149, \url{https://doi.org/10.1109/TIT.1982.1056490}.

\bibitem{zamir14book}
{\sc R.~Zamir}, {\em Lattice Coding for Signals and Networks}, Cambridge
  University Press, Cambridge, UK, 2014,
  \url{https://doi.org/10.1017/CBO9781139045520}.

\bibitem{zamir96}
{\sc R.~Zamir and M.~Feder}, {\em On lattice quantization noise}, {IEEE} Trans.
  Inf. Theory, 42 (1996), pp.~1152--1159,
  \url{https://doi.org/10.1109/18.508838}.

\end{thebibliography}

\end{document}